\documentclass{aa} 

\usepackage{amsmath}
\usepackage{soul}
\usepackage{tablefootnote}
\usepackage{bm}
\usepackage{xcolor}
\usepackage[normalem]{ulem}
\usepackage[
    range-units=single,         
    range-phrase=-,             
    separate-uncertainty=true,  
    multi-part-units=single     
] {siunitx}
\DeclareSIUnit\angstrom{Å}
\DeclareSIUnit\gauss{G}
\DeclareSIUnit\Ic{I_c}

\usepackage{graphicx}
\usepackage{txfonts}
\usepackage[colorlinks,linkcolor=blue,citecolor=blue,linktocpage=true,breaklinks, 
plainpages=false,urlcolor=blue]{hyperref}

\definecolor{iimm}{cmyk}{1.0,0.0,1.0,0.3}

\begin{document}

   \title{Solar internetwork magnetic fields: Statistical comparison between observations and MHD simulations}

   \author{E. Ebert
          \inst{1}, I. Mili\'{c}\inst{1}, J.M. Borrero \inst{1} 
          }

   \institute{Institute for Solar Physics (KIS), Georges-K\"ohler-Allee 401A, 79110 Freiburg, Germany\\
   \email{ebert@leibniz-kis.de}}

   \date{Received ; accepted }

 
  \abstract
   {Although the magnetic fields in the quiet Sun account for the majority of the magnetic energy in the solar photosphere, inferring their exact spatial distribution, origin, and evolution poses an important challenge because the signals lie at the limit of today's instrumental precision. This severely hinders and biases our interpretations, which are mostly made through nonlinear model-fitting approaches.}
   {Our goal is to directly compare simulated and observed polarization signals in the Fe I $\SI{6301}{\angstrom}$ and $\SI{6302}{\angstrom}$ spectral lines in the very quiet Sun, the so-called solar internetwork (IN). This way,  we aim to constrain the mechanism responsible for the generation of the quiet Sun magnetism while avoiding the biases that plague other diagnostic methods.}
   {We used three different three-dimensional radiative magneto-hydrodynamic simulations representing different scenarios of magnetic field generation in the internetwork: small-scale dynamo, decay of active regions, and horizontal flux emergence. We synthesized Stokes profiles at different viewing angles and degraded them according to the instrumental specifications of the spectro-polarimeter (SP) on board the Hinode satellite. Finally, we statistically compared the simulated spectra to the { Hinode/SOT/SP} observations at the appropriate viewing angles.}
   {Of the three simulations, the small-scale dynamo best reproduced the statistical properties of the observed polarization signals. This is especially prominent for the disk center viewing geometry, where the agreement is excellent. Moving toward more inclined lines of sight, the agreement worsens slightly.}
   {The agreement between the small-scale dynamo simulation and observations at the disk center suggests that small-scale dynamo action plays an important role in the generation of quiet Sun magnetism. However, the magnetic field around 50~km above the continuum layer in this simulation does not reproduce observations as well as at the very base of the photosphere.}
     \titlerunning{IN magnetic fields: comparison between observations and MHD simulations}
    \authorrunning{Ebert et al.}
   \keywords{Sun: photosphere Sun: magnetic fields}

   \maketitle


\section{Introduction}
\label{sec:intro}
 
Magnetic fields in the solar internetwork (IN) have an average magnitude of $\approx 100$\,G \citep{horizontal2, hanle}. Although much weaker than in active regions, once it is integrated over the entire solar surface, the IN makes up the majority of the solar magnetic energy. { Variation of the IN fields with solar cycle has been a topic of intensive observational study \citep{liteshinodeperiodic, fauroberpowerspectra, trellesagregor}. Other studies have focused on the inference of the orientation of the IN fields, as this also has important implications for the origins of quiet Sun magnetism \citep{steiner_simul,manfred_simul}}. This particular aspect has been investigated in recent years with a myriad of methods and data, with many different outcomes. \citet{horizontal1} and \citet{horizontal2} applied Milne-Eddington Stokes inversions to spectropolarimetric data from { Hinode/SOT/SP} of the Fe I line pair at $\SI{630}{\nano \m}$ recorded at the disk center and found a preference for horizontal fields. \citet{vertical} analyzed Sunrise/IMAX data \citep{imax} and found mostly vertical fields in the internetwork using geometrical considerations on magnetic bright spots in different layers of the atmosphere. 
\citet{isotropic} found an isotropic distribution when analyzing the two spectral lines of Fe I at $\SI{1565}{\nano \m}$. These lines are more sensitive to the horizontal component of the field and thus reduce the bias that arises from the very different sensitivities exhibited by the linear and circular polarization to the perpendicular and parallel components of the magnetic field\footnote{In this work, we refer to the components of the magnetic field that are parallel and perpendicular to the observer's line of sight as $B_\parallel$ and $B_\perp$, respectively. Other works instead use $B_{\rm LOS}$ and $B_{\rm TRA}$.}. An isotropic distribution was also favored by \citet{andres2009} who carried out a Bayesian analysis of { Hinode/SOT/SP} data. Finally, \citet{Danilovic2016_2D}, using spatially coupled inversions of Hinode/SOT/SP data, found a quasi-isotropic distribution of the magnetic field but emphasized that the distribution cannot be constrained in detail. A more detailed summary of these results can be found in \citet{borrero2017}. 

The main reason for these discrepancies is the fact that spectropolarimetric observations of the quiet Sun lie at, or even below, the noise limit of today's instruments. The photon noise limits our sensitivity to the magnetic field and its orientation \citep{photonnoise}, and cannot be increased arbitrarily without losing the temporal or spatial resolution needed to resolve the changing atmosphere. Furthermore, even in the best-case scenario, the spatial resolution of our telescopes sets a lower limit on the magnetic concentrations we can resolve using the Zeeman effect. On scales { much smaller than the resolution element}, where the magnetic field is tangled, opposite polarities in the same pixel cause the circular { polarization} to cancel out, as shown by \citet{limitedres}, making us blind to the existence of the magnetic field. {This drawback can be partially mitigated by using Hanle diagnostics \citep[e.g.,][]{hanle, tanausuavg}, but these again require specific observations and knowledge of the so-called zero-level polarization.}


To mitigate or circumvent some of the aforementioned shortcomings, we propose a statistical comparison of the observed signals in Hinode's Fe I line pair at $\SI{630}{\nano \m}$ at several heliocentric viewing angles $\Theta$ { (i.e., the angle between the line of sight and the surface normal)} to synthetic Stokes profiles calculated from three different three-dimensional radiative magnetohydrodynamic (RMHD) simulations of the solar surface convection. Before comparing with the observations, the synthetic profiles are degraded to match Hinode's spatial and spectral resolution as well as sampling. Finally, photon noise is added at the same level as in the observations. After that, histograms of the polarization signals of the observations and simulations are compared. 

On the one hand, the advantage of this approach is that we do not perform any Stokes inversions and therefore our results are not affected by the inherent issues and biases introduced by those analysis techniques \citep{photonnoise,borrero2012,borrero2013}. On the other hand, due to the limited number of simulations employed in this work, all we can aim for is to discriminate between different physical scenarios described by these simulations.

Each of the three employed simulations uses different initial conditions: small-scale dynamo, constant vertical field, and horizontal flux emergence. Our goal is to decipher which of these scenarios best reproduces the observations.{ These initial conditions} are meant to represent the most plausible origins for the IN magnetic fields \citep{luis_review}: magnetic fields amplified locally in the solar photosphere by a small-scale local dynamo mechanism \citep{originDynamo}, decay or recycling of the magnetic fields
 of active regions \citep{originDecay}, and emergence of horizontal magnetic fields from the convection zone by the global solar dynamo \citep{originDynamolarge}.

\section{Observational data}
\subsection{Observations}
\label{sec:obs}

The observations used in this work were recorded by the spectropolarimeter (SP) \citep{Spectro-Polarimeter} attached to the optical telescope \citep{SOT} on board the Hinode spacecraft \citep{Hinode}. These observations comprise the so-called Stokes vector $\Vec{I}=(I, Q, U, V),$ where $I$ is the total intensity, $Q$ and $U$ refer to the linear polarization, and $V$ is the circular polarization. The spectral coverage of{ Hinode/SOT/SP} goes from $6300.8$ to $6303.2\,\rm{\AA}$, covering the magnetically sensitive spectral lines of neutral iron: Fe I $\SI{6301.5}{\angstrom}$ and Fe I $\SI{6302.5}{\angstrom}$. This spectral region is covered by 112 wavelength points, with a wavelength sampling of 21.5 m{\AA}.

\begin{figure}
    \centering
    \includegraphics[width=0.49\textwidth]{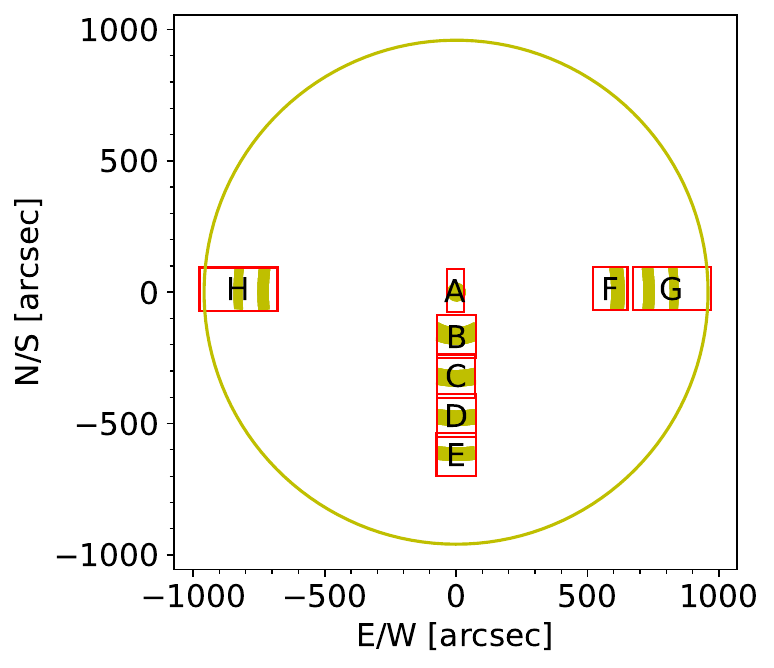}
    \caption{Position of each map on the solar disk in Table~\ref{tab:maps} as observed by the Hinode satellite. Shaded areas indicate the regions where $\Delta\Theta = \pm 2^{\circ}$ around the selected $\Theta$ values of $[0^{\circ},10^{\circ},20^{\circ},30^{\circ},40^{\circ},50^{\circ},60^{\circ}]$ (see text for details).\label{fig:maps_on_disc}}
\end{figure}

We chose eight maps with different heliocentric positions, which are listed in Table~\ref{tab:maps}. Maps B to H are a part of the observational table compiled by \citet{2017ApJ...835...14L}. Figure~\ref{fig:maps_on_disc} shows the locations of the maps on the solar disk. The field of view along the slit in the north--south direction is $\SI{164}{\arcsecond}$,  which is sampled in steps of $\SI{0.16}{\arcsecond}$ (i.e., equal to the width of the slit). This sampling results from the angular diffraction limit $DL = 1.22 \frac{\lambda}{D} = \SI{0.32}{\arcsecond}$, where $D = \SI{0.5}{\m}$ is the diameter of the Hinode telescope. This results in a pixel size, at disk center, of $d_{s} \cdot \sin{\SI{0.16}{\arcsecond}} = \SI{116.4}{\km}$, where $d_s$ is the average Earth--Sun distance. The reason for choosing these specific maps is that the observations were all performed within one week, and so we can neglect changes in the solar atmosphere on longer timescales. In this period, no significant degradation of the instrumentation is expected. All the maps have the same total integration time per slit scan position of $\Delta T = \SI{9.6}{s}$ resulting in a similar noise level.

\begin{figure}
    \centering
    \includegraphics[width=0.49\textwidth]{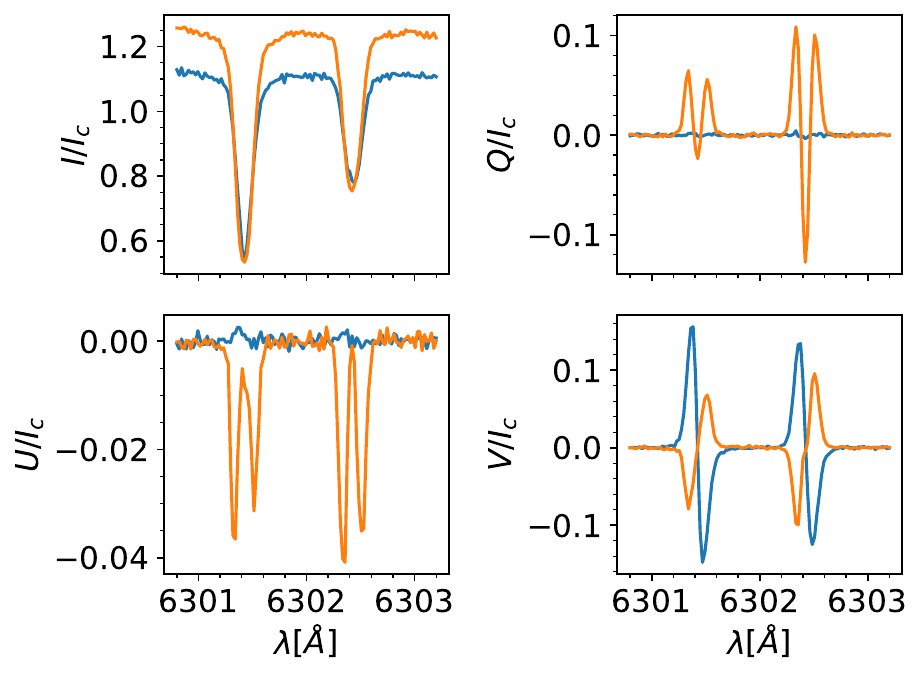}
    \caption{Stokes profiles of a pixel at { $\Theta = 0^\circ$ (blue) and $60^\circ$ (orange)}. They are normalized to the mean continuum intensity at the respective heliocentric angle. \label{fig:profile}}
\end{figure}

\begin{table*}
    \centering
    \caption{Observational maps used for the comparison}
    \begin{tabular}{c c c c c c c c c}
         \hline \hline
         Map & $\Theta^a$ &  Date &  UT & Xcen$^b$ & Ycen$^c$  & $\sigma_Q^d$ [$\times \num{ e-4}]$ & $\sigma_U^e$ [$\times \num{ e-4}$] & $\sigma _V^f$ ’ [$\times \num{ e-4}]$ \\
         \hline
         A & \SI{0}{\degree}  & 2007 Sep 10 & 08:00 &-3 & 7    & 8.3 & 8.2 & 7.7 \\
         B & \SI{10}{\degree} & 2007 Sep 15 & 12:44 & 3 & -196 & 8.3 & 8.2 & 7.8 \\
         C & \SI{20}{\degree} & 2007 Sep 15 & 15:34 & 1 & -319 & 8.3 & 8.3 & 7.8 \\
         D & \SI{30}{\degree} & 2007 Sep 16 & 12:31 & 1 & -469 & 8.6 & 8.5 & 8.0 \\
         E & \SI{40}{\degree} & 2007 Sep 16 & 15:34 & 0 & -619 & 9.0 & 8.9 & 8.4 \\
         F & \SI{40}{\degree} & 2007 Sep 08 & 15:34 & 587 & 13 & 9.0 & 8.9 & 8.4 \\
         G & \SI{50}{\degree} & 2007 Sep 09 & 07:05 & 808 & 14 & 9.5 & 9.4 & 9.0 \\
         G & \SI{60}{\degree} & 2007 Sep 09 & 07:05 & 808 & 14 & 10.0 & 10.0 & 9.6 \\
         H & \SI{50}{\degree} & 2007 Sep 06 & 07:04 &-798 & 12 & 9.6 & 9.6 & 8.9 \\
         H & \SI{60}{\degree} & 2007 Sep 06 & 07:04 &-798 & 12 & 10.4 & 10.3 & 9.6 \\
         \hline
    \end{tabular}
    \label{tab:maps}
    \tablefoot{
        \tablefoottext{a} {Map average viewing angle $\Theta$}        
        \tablefoottext{b} {arcseconds east of disk center to map center}
        \tablefoottext{c} {arcseconds north of disk center to map center}
        \tablefoottext{d} {inferred noise of $Q$}
        \tablefoottext{e} {inferred noise of $U$}
        \tablefoottext{f} {inferred noise of $V$}
    }
\end{table*}

\subsection{Preprocess}
\label{Section::Obs_preprocess}

The value of the heliocentric angle $\Theta$ for the central position of each map was determined by extracting the heliocentric $(x_c, y_c)$ coordinates from the header of each FITS file and applying:

\begin{equation}
    \label{eq:mu_from_decl_RA}
    \Theta = \arcsin{\left(\frac{d_{s}}{R_{Sun}\sqrt{x_c^2+y_c^2}}\right)}
.\end{equation}
 
We note that, according to \citet{inacXCEN}, the pointing information of Hinode is inaccurate by $\Delta x_c \approx \Delta y_c \approx \pm \SI{30}{\arcsecond}$. This means that the value of $\Theta$ determined by Eq.~\ref{eq:mu_from_decl_RA} is subject to an error, referred to as $\sigma_\Theta$, and given by:

\begin{equation}
    \label{eq:delta_mu_from_decl_RA}
    \sigma_\Theta \approx \Delta x_c \frac{d_{s}}{\sqrt{R_{Sun}^2-d_s^2[x_c^2+y_c^2]}}
.\end{equation}

Equation~\ref{eq:delta_mu_from_decl_RA} results in an error of $\sigma_\Theta \approx \SI{1.8}{\degree}$ at $\Theta = \SI{0}{\degree}$ and this increases to $\sigma_\Theta \approx \SI{3.6}{\degree}$ at $\Theta = \SI{60}{\degree}$. This is important because for Map H, which contains the solar limb, the $x_c$ value provided by the headers had to be corrected by $\SI{+3}{\arcsecond}$ such that the limb was at $\SI{959}{\arcsecond}$. For map G, also containing the limb, the  $x_c$ value provided by the headers had to be corrected by $\SI{+7}{\arcsecond}$.

Now, each of the selected maps spans a significant portion of the solar disk and therefore has a different viewing angle $\Theta$ for different pixels. The difference between the maximum and minimum value of $\Theta$ for a given map is larger for the observations closer to the solar limb. Therefore, to avoid mixing observations from pixels with very different viewing angles contained in the same map, we restrict our analysis to those pixels within each map where the viewing angle is within $\pm 2^{\circ}$ of the following nominal values: $\Theta \in [\SI{0}{\degree},\SI{10}{\degree},\SI{20}{\degree},\SI{30}{\degree},\SI{40}{\degree},\SI{50}{\degree},\SI{60}{\degree}]$. These restricted regions are shown as shaded areas  in Fig.~\ref{fig:maps_on_disc} for each of the eight selected maps.

Every quiet Sun map was normalized to its local continuum intensity (using only the selected pixels within $\pm 2^{\circ}$) by dividing all four Stokes components for all wavelengths by the mean of the local continuum intensity. The noise in the measurement of the polarization signals ($Q, U, V$) is assumed to have a Gaussian distribution as they are obtained from the difference between two simultaneous intensity measurements. This cancels out any systematic error in the intensity measurement that might have come from imperfect flat-field correction, dark current, and so on. The level of photon noise is different for different Stokes components (see Table~\ref{tab:maps}), with the linear polarization signals ($Q$, $U$) having slightly larger values than the circular polarization signals ($V$).

At continuum wavelengths, one expects no signal in the polarization, and so we infer the noise by taking the standard deviation of the continuum polarization of all selected spatial pixels. For this, the wavelength region $\in [6300.907, 6301.121]$~{\AA} was chosen, which is expected to only contain the continuum. The inferred noise, in units of local quiet Sun continuum, is listed in Table \ref{tab:maps} as $\sigma = \frac{1}{\mathrm{S/N}}$. The signal-to-noise ratio (S/N) decreases with $\Theta$ because of the limb darkening. Namely, the noise scales with the number of photons as $\sqrt{N}$, where $N$ is the number of detected photons, and thus, S/N also scales with the intensity as $\sqrt{N}$.

{ In Fig.~\ref{fig:profile} we show example profiles at $\Theta = 0 ^\circ$ in blue and $\Theta = 60 ^\circ$ { in orange}. As opposed to randomly choosing two pixels, which would likely feature mostly noise, the specific pixels were chosen because of their strong polarization signals. The pixel at $\Theta = 60 ^\circ$ has also a strong signal in both linear polarizations to highlight the fact that the linear polarization is stronger when viewing from a higher inclination to the solar surface. Both pixels have such a high circular polarization that they will be excluded from the analysis as they will be considered to be part of the solar magnetic network (as explained in \ref{sec:: removal of network})}

\begin{figure}
    \centering
    \includegraphics[width=0.5\textwidth]{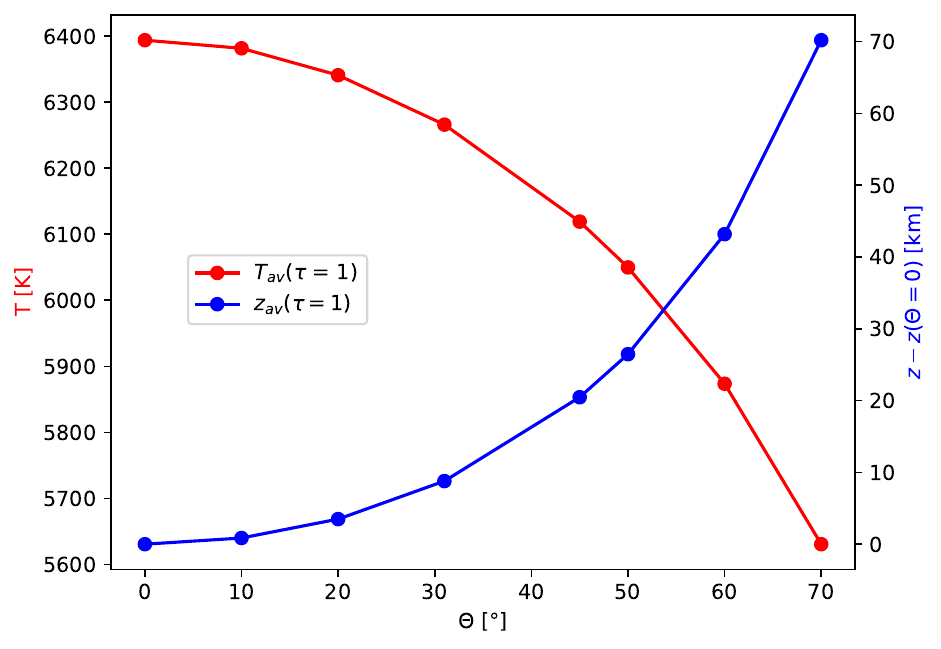}
    \caption{Average geometrical height and average temperature at $\tau = 1$ along the line of sight for simulation \#2 (see Sect.~\ref{sec:simulations}) as a function of heliocentric angle $\Theta$.\label{fig:zoftau}}
\end{figure}

For completeness, we provide the height of the optical depth $\tau=1$ along the line of sight for different viewing angles in Fig.~\ref{fig:zoftau} (blue). As we can see, $z(\tau=1)$ increases as we observe closer to the limb. This increase is accompanied by a decrease in the average temperature at the observed $\tau=1$ surface and a decrease in the observed intensity (i.e., limb darkening).

\section{Synthesis of simulations and degradation}
\label{sec:simulations}
\subsection{Description of simulations}

The first two simulations used in this paper were performed using the MURaM radiative MHD code \citep{MURAM}. These simulations have a { spatial} domain of $24.576 \ \times \ 24.576 \ \times \ 7.680 \ \SI{}{\mega\m^3}$. The top boundary is located at about $\SI{1.5}{\mega\m}$ above the average $\tau = 1$ layer. This leaves a depth for the convective part of about $\SI{6.2}{\mega\m}$. For the synthesis, only the upper $\SI{2.048}{\mega \m}$ are used, as lower depths do not contribute to the emergent intensity and a lot of computing time can be saved. These simulations have a sampling of $\Delta x = \Delta y = \Delta z = \SI{16}{\km}$ in all spatial dimensions, with a total number of $n_x = n_y = 1536$ and $n_z = 480$ grid cells.

Simulation \#1 is a snapshot directly taken from Z16M in \citet{Numerical-sim}. This simulation uses boundary conditions that are antisymmetric in inflow regions ($\Vec{B}$ is set to zero) and symmetric in outflow regions. Here, the magnetic field arises from the small-scale dynamo, meaning that the simulation starts from a thermally relaxed convection simulation with no magnetic field to which a weak ``seed field'' of  $\SI{0.005}{\gauss}$ is added. Then the simulation evolves in time while the dynamo enhances the magnetic field until it reaches a saturated phase. This simulation represents a scenario where the small-scale local dynamo (SSD) is the sole origin of the IN fields \citep{originDynamo}.

Simulation \#2 was obtained by taking a snapshot from a similar small-scale dynamo simulation \citep[referred to as O16bM in][]{Numerical-sim} and then adding a constant $B_z=\SI{30}{\gauss}$ throughout the entire box and letting it evolve for 6 additional hours of solar time. Here the boundary conditions prescribe a symmetric behavior for all physical parameters including the magnetic field along the bottom. Simulation \# 2 is designed to represent a scenario where the IN fields originate from decaying fields of active regions \citep{originDecay}.

Simulation \#3 was carried out by \citet{PhDFlavio} ---where it is referred to as d3gt57g44h50mfc--- using the magneto-hydrodynamic CO5BOLD Code \citep{Cobold}. It covers a horizontal section of $9.6 \ \times \ 9.6 \ \SI{}{\mega\m^2}$ of the solar atmosphere and extends from the top of the convection zone to beyond the top of the photosphere over a height range of $\SI{2.8}{\mega \m}$, which is about $\SI{1240}{\km}$ below and $\SI{1560}{\km}$ above the solar surface ($\tau = 1$). It has a sampling of $\Delta x = \Delta y = \Delta z = \SI{10}{\km}$ in all spatial dimensions, with a total number of $n_x = n_y = 960$ and $n_z = 280$  grid cells. The initial magnetic field is zero everywhere and a horizontal magnetic field with an absolute flux density of $\SI{50}{\gauss}$ is advected through the bottom boundary. At the top boundary, the magnetic field is forced to be vertical in downflow regions and is constantly extrapolated in upflow regions. The simulation runs for $\SI{4.1}{\hour}$, after which we choose the snapshot for the present analysis. This simulation represents a scenario where the{ IN magnetism} originates from the appearance of magnetic fields ---initially located{ in} the convection zone--- at the solar surface, as part of the global dynamo \citep{originDynamolarge}.

The mean unsigned and mean signed vertical magnetic field, and the mean total magnetic field, in the simulations at $\tau = 1$ of the selected snapshots of each simulation are shown in Table~\ref{tab:B}. We note that the mean signed vertical magnetic field is not exactly 0~G for simulations \# 1 and \#3, nor is it exactly 30~G for simulation \# 2, because the values in this table correspond to a constant optical depth $\tau$ rather than to a constant height $z$.

\begin{figure}
    \centering
            \includegraphics[width=0.5\textwidth]{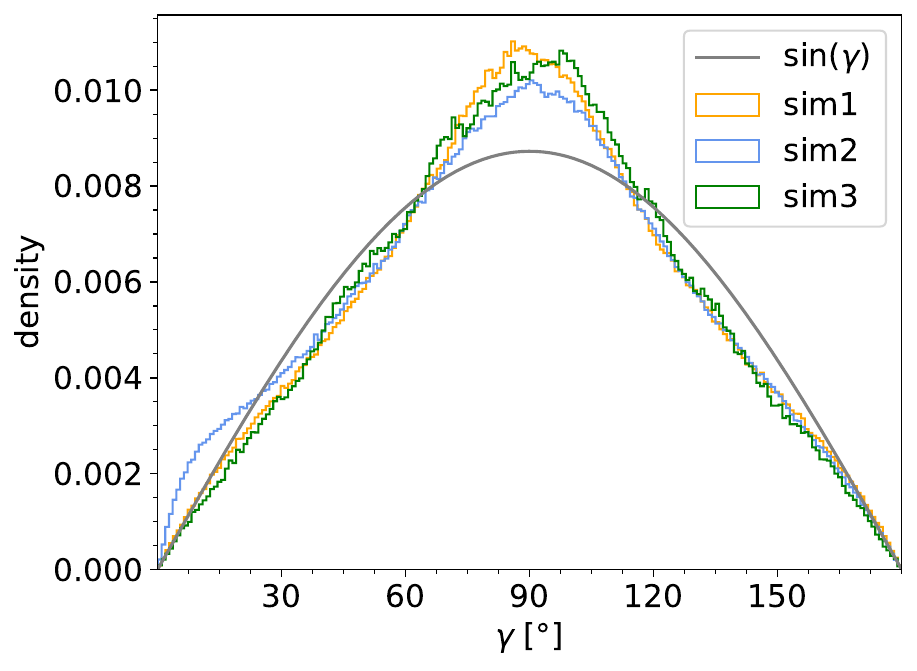}
            \caption{Histogram of the inclination of the magnetic field at $\tau = 1$,  where $\sin{\gamma}$ ({ gray} line) corresponds to an isotropic distribution\label{fig:sim_inclination}.}
\end{figure}

Owing to our interest in investigating mainly the angular distribution of the magnetic field in the IN (see Sect.~\ref{sec:intro}), we compare, in Fig.~\ref{fig:sim_inclination}, the distribution of the inclination of the magnetic field with respect to the observer's line of sight ($\gamma$) at $\tau = 1$ from all three simulations. One can see that they all favor horizontal fields because there is an excess of pixels with respect to the isotropic distribution (gray line), with an inclination of between $\in [60^{\circ},120^{\circ}]$. As demonstrated by \citet{borrero2013}, an isotropic distribution follows a{ sinusoidal} function that comes from the Jacobian when transforming to spherical coordinates. For simulation \#2, the vertical magnetic field imposed as an initial condition can be seen at the $\tau=1$ level as an overabundance of pixels with an inclination of $\gamma \in [\SI{0}{\degree},\SI{30}{\degree}]$ (blue line).

\begin{table}[h]
    \centering
    \caption{Magnetic field in the three simulations}
    \begin{tabular}{c c c c}
    \hline
        simulation & $\langle |B_z| \rangle$ [$\SI{}{\gauss}$] & $\langle B_z \rangle$ [$\SI{}{\gauss}$] & $\langle B\rangle$ [$\SI{}{\gauss}$]\\
    \hline
        1 & $66.2$ & $-0.1$ & 125.0\\
        2 & $89.1$ & $28.7$ & 149.4\\
        3 & $61.1$ & $-0.4$ & 125.3\\      
    \end{tabular}
    \tablefoot{Mean unsigned vertical component of the magnetic field { (second column)}, mean signed vertical component of the magnetic field { (third column)}, and mean total vertical magnetic field ({fourth column)} in the simulations. All values correspond to $\tau = 1$.}
    \label{tab:B}
    
\end{table}

\subsection{Synthesis}
\label{sec:synthesis}

To compare the predictions from the three simulations with observations, the emergent polarized spectra (i.e., the Stokes profiles) were synthesized from the simulation snapshots (i.e., MHD cubes) using the code FIRTEZ-dz \citep{Firtez}. We solve the radiative transfer equation for polarized light under the Zeeman effect, assuming local thermodynamic equilibrium (LTE) to solve the equation of state and find the populations of the atomic levels of the two lines observed by { Hinode/SOT/SP} (see Sect.~\ref{sec:obs}). The parameters of the spectral lines are shown in Table~\ref{tab:line_para}. The data were synthesized with the same sampling as the observations of $\SI{21.5}{\milli \angstrom}$. 

\begin{table*}
    \centering
    \caption{Atomic parameters of the neutral iron lines used for this study.}
    \begin{tabular}{c c c c c c c c c}
         \hline \hline
         Element & Ion &  $\lambda_0 [\SI{}{\angstrom}]$$^a$ &  log(gf)$^b$ & $\chi_{\mathrm{low}}$$^c$ [eV] & $\alpha$$^d$ & $\sigma/(a_0^2)$$^e$  & Upper$^f$ & Lower$^f$ \\
         \hline
         Fe & I & 6301.5012 & -0.718 & 3654 & 0.243 & 840 & $^5P_2$ & $^5D_2$ \\
         Fe & I & 6302.4936 & -1.165 & 3686 & 0.241 & 856 &$ ^5P_1$ & $^5D_0$ \\
         \hline
    \end{tabular}
    \label{tab:line_para}
    \tablefoot{
        \tablefoottext{a} {Central wavelength $\lambda_0$}        
        \tablefoottext{b} {oscillator strength}
        \tablefoottext{c} {excitation potential of the lower level $\chi_\mathrm{low}$}
        \tablefoottext{d} {temperature exponent of the collisional broadening under the ABO theory}
        \tablefoottext{e} {cross section, in units of the square of the Bohr radius $a_0^2$, of the collisional broadening under the ABO theory \citep{anstee1995}}
        \tablefoottext{f} {electronic configurations of the lower and upper levels}
        the table values are taken from \citet{Fistro2014} but corrected for the typo of the electronic configuration of the lower level of the $\SI{6301}{\angstrom}$ line}
\end{table*}

To calculate the spectra for viewing angles $\Theta>0$, FIRTEZ-dz rotates the simulated atmosphere by a given angle and uses trilinear interpolation to determine the physical parameters along the inclined ray (i.e., on a new 3D grid). The essential aspect of this rotation is the assumption of periodic boundary conditions (also used in all three simulations), as it allows us to preserve the spatial extent of our boxes when doing the rotation.

 \begin{figure*}
    \begin{center}
    \begin{tabular}{cc}
         \includegraphics[width=8cm]{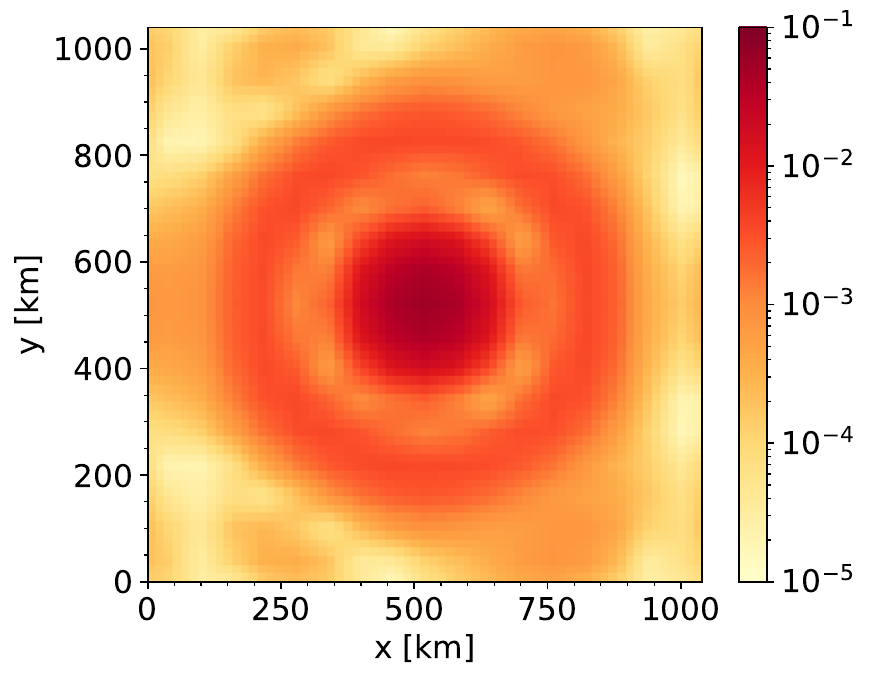} &
         \includegraphics[width=8.1cm]{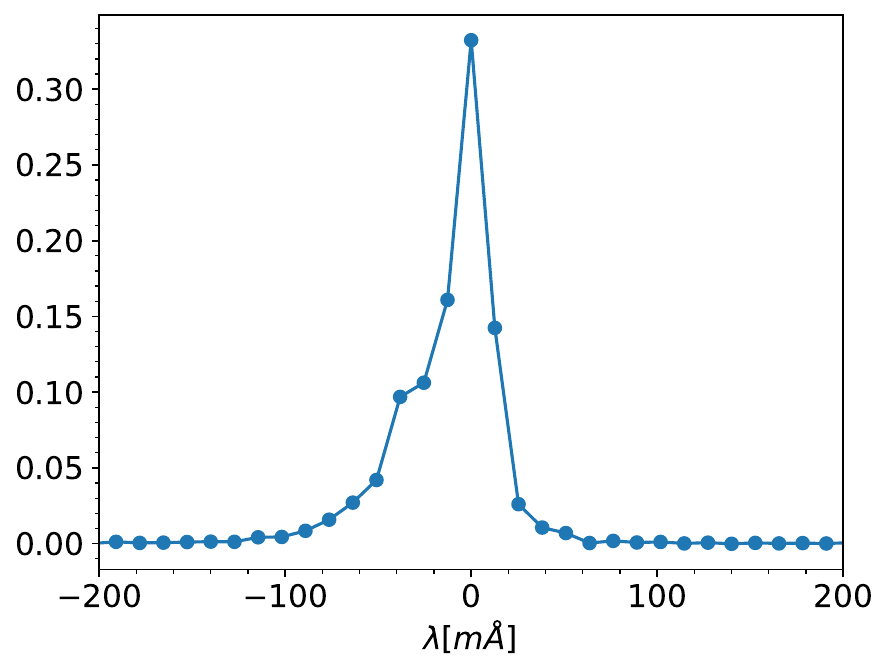}
    \end{tabular}
     \caption{Spatial and spectral degradation of the simulated Stokes profiles. {Left panel}: Spatial PSF of the { Hinode/SOT/SP} instrument (at disk center). {Right panel}: { Hinode/SOT/SP} spectral transmission curve.\label{fig:spatialPSF}}
     \end{center}
 \end{figure*}
 
\subsection{Degradation}
\label{sec:degradation}

To make the simulated data comparable to the observed data, they need to be degraded to fit the spectral and spatial resolution of {the Hinode/SOT/SP} instrument. We used a spectral transmission profile (provided courtesy of Lites B.W. and shown in the right panel of Fig.~\ref{fig:spatialPSF}) to spectrally convolve the synthetic data, thus effectively degrading its spectral resolution to $\approx \SI{30}{\milli\angstrom}$. In addition, we used the spatial point-spread function (PSF) taken from \citet{contrast} to degrade the spatial resolution to $\SI{0.16}{\arcsecond}$. The spatial PSF at the disk center ($\Theta=0^{\circ}$) is also shown in Fig. \ref{fig:spatialPSF} (left panel). For other $\Theta$ values, the spatial PSF is modified to account for the larger pixel size along one spatial dimension. Finally, we resampled the data in $x$ and $y$ directions to fit the{ Hinode/SOT/SP} pixel size using Eq. \ref{eq:resampling}, which gives the new number of grid cells, $n_x^{\prime}$ and $n_y^{\prime}$, along each of the horizontal dimensions:

\begin{eqnarray}
\label{eq:resampling}
    n_{x}^{\prime} & = & n_{x} \cos\Theta \frac{\Delta x}{d_s \sin(\SI{0.16}{\arcsecond})} \notag\\
    n_{y}^{\prime} & = & n_{y} \frac{\Delta y}{d_s \sin(\SI{0.16}{\arcsecond})} \;,
\end{eqnarray}

\noindent where the foreshortening according to the viewing angle $\Theta$ is only done along one of the spatial dimensions. For instance, the above equations indicate that the simulation \#1, which originally had $n_x=n_y=1536,$ has to be resampled into $n_x^{\prime} = n_y^{\prime} = 212$ grid cells at $\Theta=0^{\circ}$, but into $n_x^{\prime} = 183$ and $n_y^{\prime} = 212$ grid cells at $\Theta = 30^{\circ}$. The last step before comparing the simulated data with the observed data is to add the same noise level as was obtained for the observations in Sect.~\ref{Section::Obs_preprocess}. The noise vector $\Vec{n}_\Vec{I}$ was created by drawing from a Gaussian distribution with standard deviations $\sigma_Q$, $\sigma_U$,  or $\sigma_V$ (see Table~\ref{tab:maps}). We used wavelength-dependent noise so that the new polarization signals with noise added are given as: 

\begin{equation}
\Vec{I}_{\rm new}(x,y,\lambda) = \Vec{I}_{old}(x,y,\lambda) + \sqrt{I_{\rm old}(x,y,\lambda)} \Vec{n}_\Vec{I}(\lambda) \;,
\label{eq:noise}
\end{equation}

\noindent where $I_{\rm old}(x,y,\lambda)$ is the normalized intensity profile. This ensures that less noise is added closer to the line core than in the continuum because the number of detected photons is lower. Equation~\ref{eq:noise} applies to both the intensity ($I_{\rm new}$) and the polarization ($Q_{\rm new}$, $U_{\rm new}$, and $V_{\rm new}$) because the noise in the polarization profiles is the same as in the intensity, including its wavelength dependence. This happens because the latter is obtained from the linear combination of intensity measurements \citep[i.e., modulation; ][]{demodulation}.

\subsection{Removal of network}
\label{sec:: removal of network}

\begin{figure*}
\begin{tabular}{ccc}
    \includegraphics[width=0.33\textwidth]{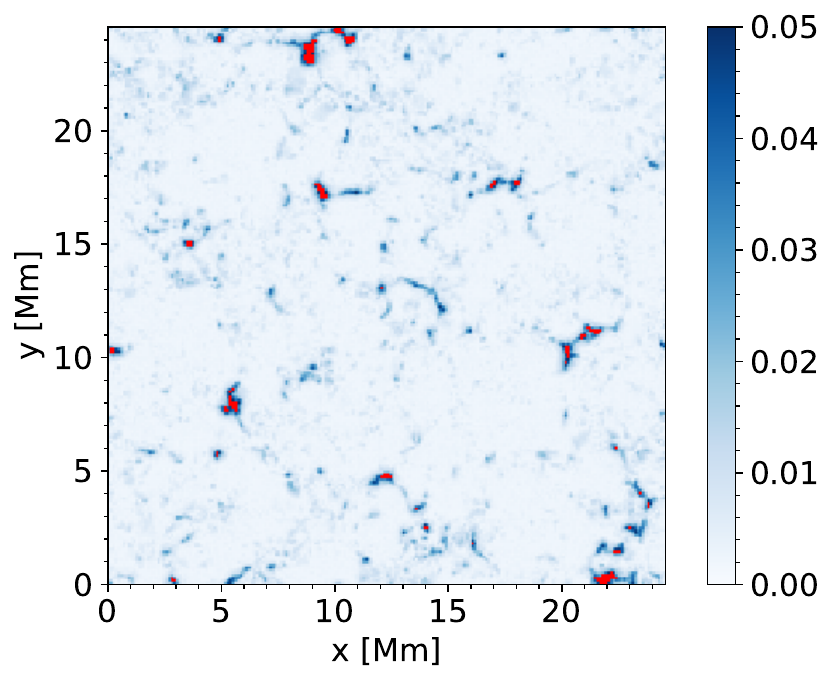} &
    \includegraphics[width=0.33\textwidth]{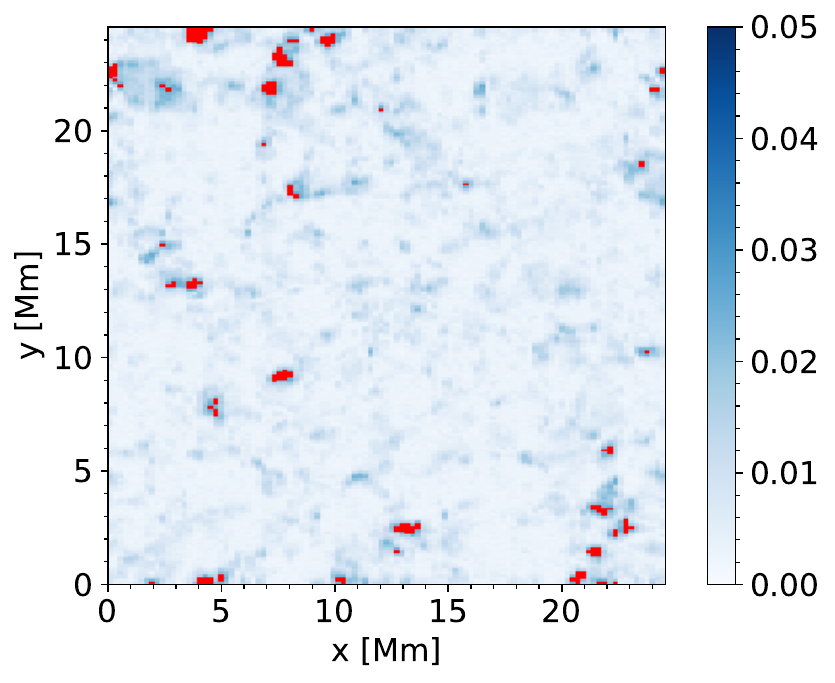}&
    \includegraphics[width=0.32\textwidth,trim= 0 -4 0 0, clip]{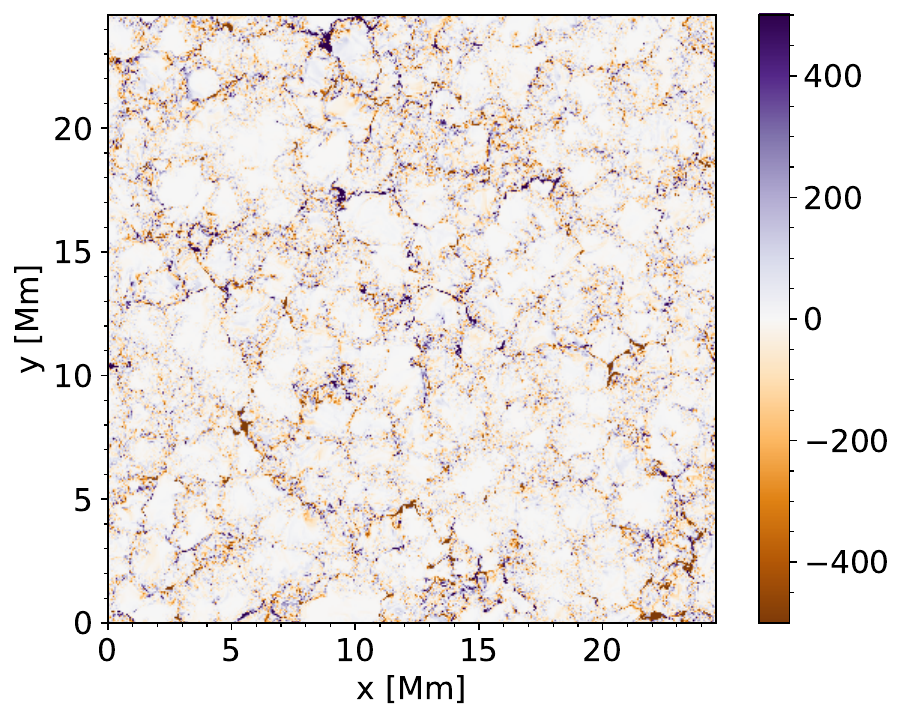}
\end{tabular}
\caption{Comparison of $\max\|V(\lambda)\|$ { in units of continuum intensity} for $\Theta = \SI{0}{\degree}$ (left panel) and $\Theta = \SI{60}{\degree}$ (middle panel) with $B_z ~[G]$ of simulation \#1 (right panel).} 
\label{fig:sim2-maxV-comp} 
\end{figure*}

Our work focuses on the IN fields, and so it is necessary to remove the network regions, both from the observed and synthetic data. The network is assumed to contain strong magnetic fields ($> \SI{500}{\gauss}$), which are observed at the disk center as mostly vertical. They are also found mostly in between supergranules, at scales of $\SI{30}{\mega \m}$ \citep{scale_network}. One possibility to identify the network pixels in the observations would be to perform an inversion of the Stokes profiles and look for regions where the magnetic field features these network-like properties. However, our study avoids Stokes inversions to circumvent the biases described by \cite{borrero2013}. Thus, we need to identify network pixels directly from the observed Stokes vector instead of using magnetic field values from the inversion. Although the horizontal size in our simulations is smaller than the spatial scales of the network \citep[i.e., 9.6-24.576~Mm vs 30~Mm; ][]{scale_network}, they do contain a few network-like patches that can be seen in the vertical magnetic field (see the rightmost panel in Fig.~\ref{fig:sim2-maxV-comp}). These patches of strong vertical magnetic field were then considered to be{ part of the} network.

As the magnetic field of the network is mostly vertical at disk center, there is a large signal in Stokes $V$. To identify the network in circular polarization, we compared in Fig.~\ref{fig:sim2-maxV-comp} the maximum of $\|V(\lambda)\|$ at $\Theta=0^{\circ}$ (left) with $B_z(x,y)$ at $\tau = 1$ (right). We determined a threshold of $\max\|V(\lambda\|) = 0.05$, above which a pixel is ascribed to the network and thus removed from our analysis.

\begin{figure}
    \centering
    \includegraphics[width=0.5\textwidth]{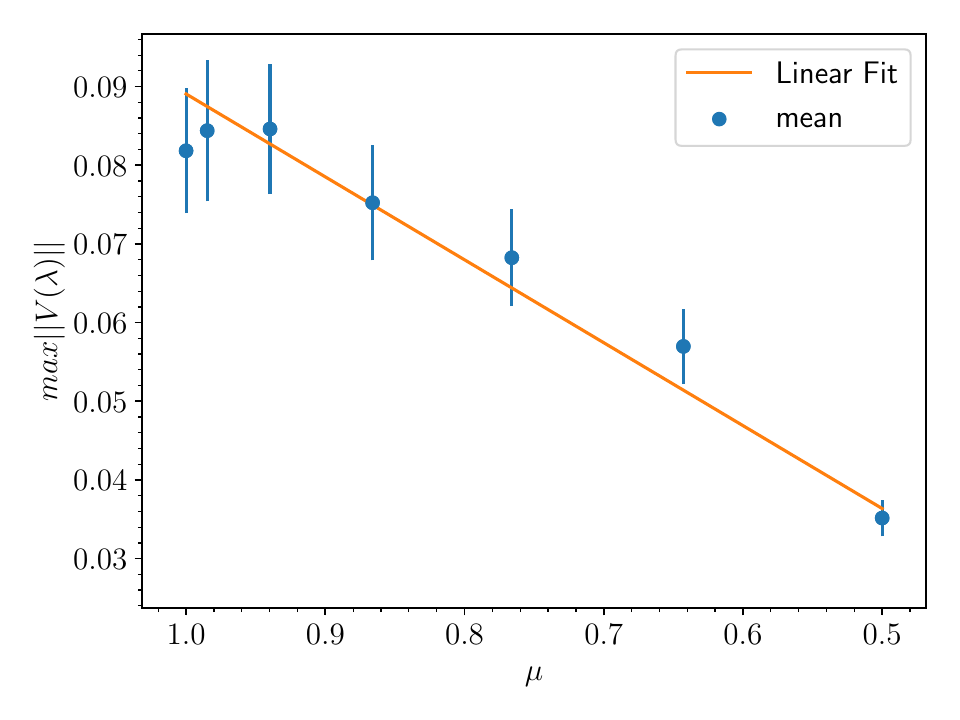}
    \caption{Empirical relationship between the $\max(|V(\lambda)|)$ and $\mu=\cos\Theta$ for the several network patches in Fig.~\ref{fig:sim2-maxV-comp}. Vertical blue bars indicate the standard deviation for the sample of about 500 pixels included in these patches.}
    \label{fig:network-detec}
\end{figure}

Finding a threshold to identify the network off-disk center ($\Theta>0$) requires additional considerations. Although the network is not completely homogeneous and is known for having internal structure \citep{marian_network}, we surmise that, at disk center, the network is mostly vertical. Under this assumption, the observed circular polarization or Stokes $V$ in the network should scale as $V\cos\Theta$. To test this hypothesis, we took several patches in the simulations at disk center $\Theta=0^{\circ}$ (see left panel in Fig.~\ref{fig:sim2-maxV-comp}) that were identified as network and plotted the average over those selected pixels of ${\rm max}\|V(\lambda)\|$ as a function of $\mu=\cos\Theta$. These patches include about 500 pixels. Results are presented in Fig.~\ref{fig:network-detec}. Here we can see that the circular polarization signals in the simulated network roughly scale linearly with $\cos\Theta$ (see orange curve), thereby justifying our initial assumption that network fields are mostly vertical. With this, we can now find a threshold with which to detect network patches in the simulations with $\Theta > 0$ as those pixels where ${\rm max}\|V(\lambda)\| > 0.05 \cos\Theta$. As already mentioned, pixels identified as such are masked out from our analysis. Two examples of the masked pixels in simulation \# 1 are presented in red in Fig. \ref{fig:sim2-maxV-comp} ($\Theta=0^{\circ}$ and $\Theta=60^{\circ}$ in the left and middle{ panels, respectively}).

\section{Results}
\label{sec:results}

To compare the observed{ Stokes profiles with those} synthesized from the simulations, we took a statistical approach. For each spatial pixel $(x,y),$ the maximum of the absolute value of $[Q(\lambda),U(\lambda)]$ and $V(\lambda)$ was used to make a histogram. $Q$ and $U$ are combined to form an approximate total linear polarization $L$ for each pixel by taking whichever is the largest of the two. As an example, in Figs.~\ref{fig:sim2comparisonB}
and ~\ref{fig:sim3comparisonB}  we display these histograms (in black lines) for the observed signals at $\Theta=10^{\circ}$  (map B; see Table~\ref{tab:maps}). Circular and linear polarization signals are presented in the left and right panels of these two figures, respectively.

For most of the pixels, the polarization signal is below the noise level. This can be seen in the peak of the histograms at $\approx 2.6 \cdot \sigma \approx 2.3 \times 10^{-3}$. As there are 112 wavelength pixels (see Sect.~\ref{sec:obs}), one expects the peak of these histograms to be at $\sqrt{2} \cdot \mathrm{erf^{-1}}(1-1/112) \cdot \sigma$ (where $\mathrm{erf^{-1}}$ is the inverse error function) if there is no signal in addition to the noise. Hereafter, we refer to this peak produced by photon noise simply as the noise-peak. Signals with a level above the noise peak until the drop above 0.05 are considered to be representative of the IN. The drop above the 0.05 level is caused by the removal of the network described in Sect.~\ref{sec:: removal of network}. 

{ For a quantitative comparison between the observed and simulated histograms, we performed a $\chi^2$ analysis, where the merit function $\chi^{2}$ is defined as:
 
\begin{equation}
    \chi^2 = \sum_{i=1}^k (S_i-O_i)^2/O_i
,\end{equation}

where $S_i$ and $O_i$ are the fractions of pixels in the simulation and observations, respectively, falling in between two logarithmically spaced Stokes signals as shown in Figs.~\ref{fig:sim2comparisonB} through~\ref{fig:discussion2}. This merit function was chosen to sample in $k=125$ histogram bins. A smaller $\chi^2$ value indicates a better fit between observations and simulations, but we emphasize that the values of $\chi^{2}$ can be influenced by slightly different positions of the noise peak in the observations and simulations, as well as by the existence of outliers in the simulations in bins where $O_i \rightarrow 0$. Consequently, we only draw conclusions when find agreement between a quantitative analysis in terms of $\chi^{2}$ and a qualitative analysis based on our visual perception of the similarities between simulations and observations. The specific values of $\chi^{2}$ for both the circular and linear polarization, as well as the percentage of network pixels removed, are provided in the legend of{ Figs~\ref{fig:discussion1}-\ref{fig:discussion2}.}}

\subsection{Field strength reduction in simulations \# 2 and \# 3}

\begin{figure*}
    \centering
    \includegraphics[height=0.3\textwidth]{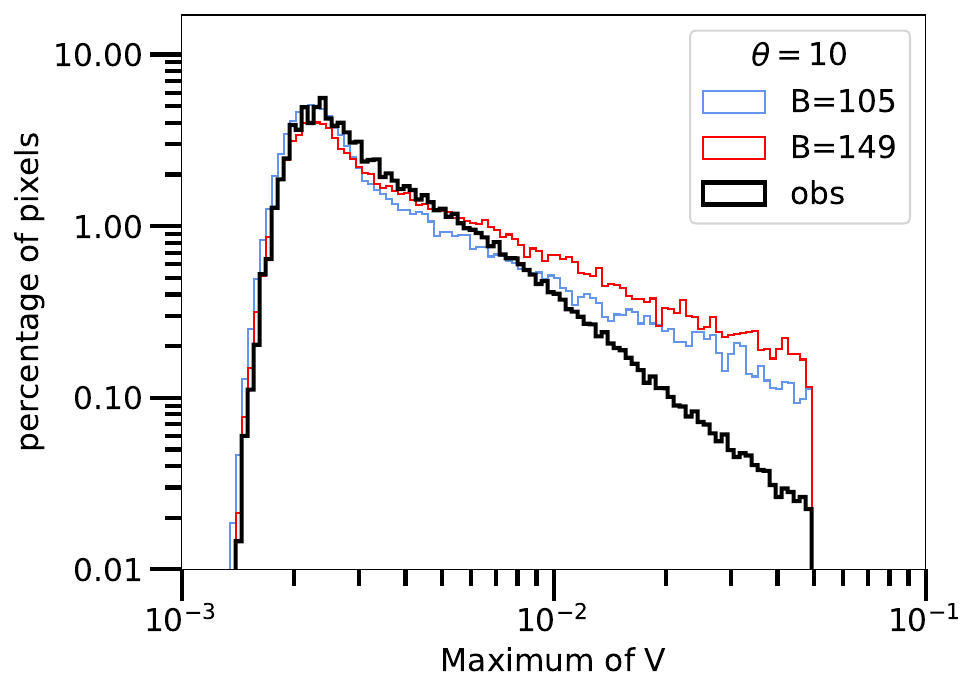}
    \includegraphics[height=0.3\textwidth]{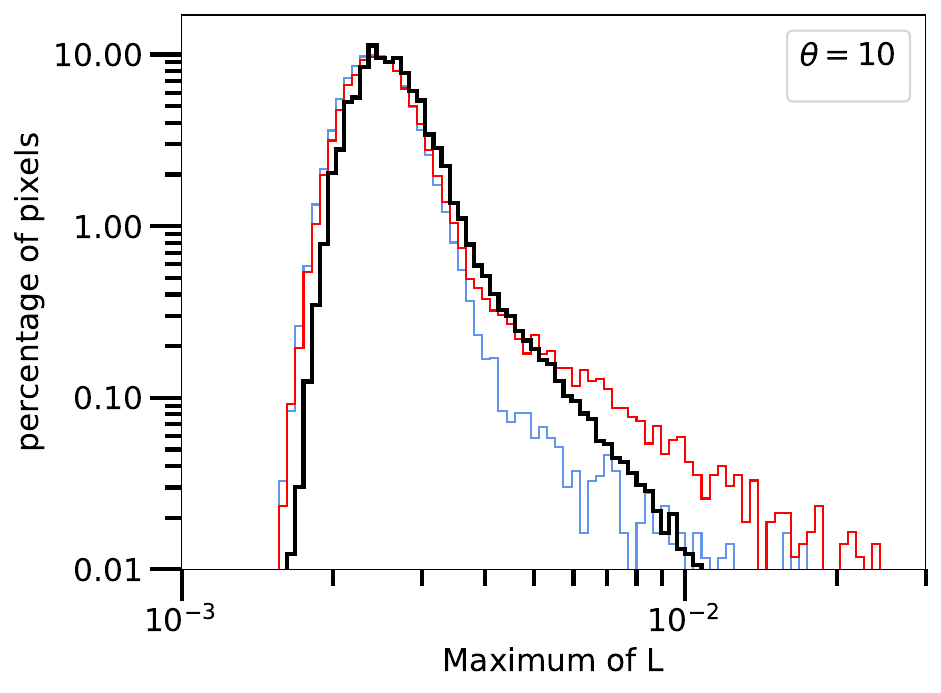}
    \caption{Histograms of the polarization signals at $\Theta=10^{\circ}$. {\it Left panel}: ${\rm max} \|V(\lambda)\|$. {\it Right panel}: ${\rm max} [\|Q(\lambda) \|,\|U(\lambda)\|$. Observed values are shown in black, whereas predicted histograms from simulation \# 2 are shown in colors for two different values of $\langle B(\tau=1) \rangle$: 149~G (red; original value) and 105~G (blue; 70\% of the original value).\label{fig:sim2comparisonB}}
\end{figure*}

\begin{figure*}
    \centering
    \includegraphics[height=0.3\textwidth]{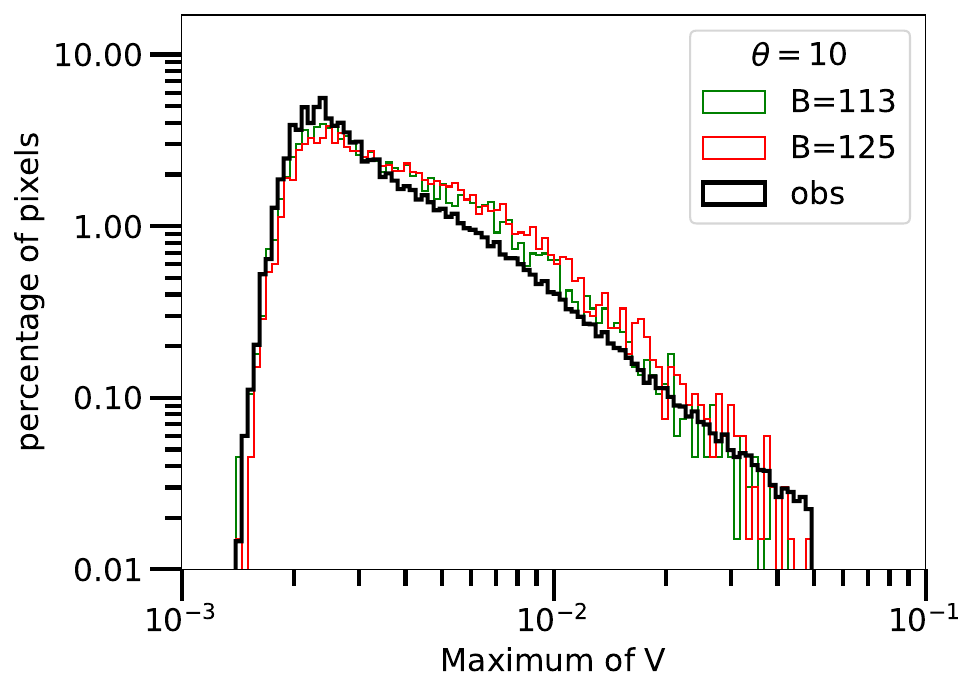}
    \includegraphics[height=0.3\textwidth]{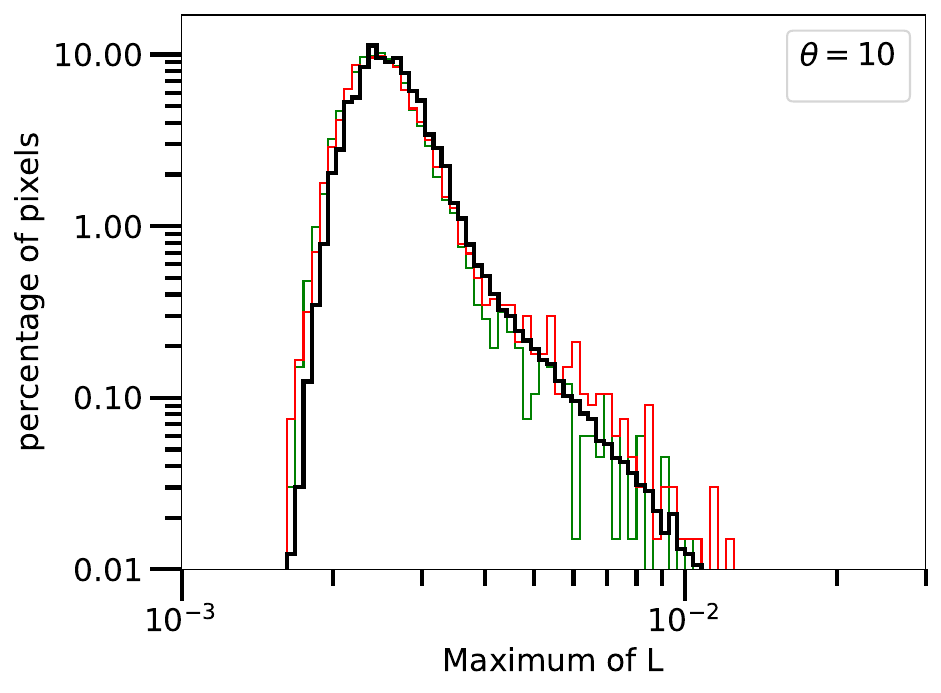}
    \caption{Histograms of the polarization signals at $\Theta=10^{\circ}$. {\it Left panel}: ${\rm max} \|V(\lambda)\|$. {\it Right panel}: ${\rm max} [\|Q(\lambda) \|,\|U(\lambda)\|$. Observed values are shown in black, whereas predicted histograms from simulation \# 3 are shown in colors for two different values of $\langle B(\tau=1) \rangle$: 125~G (red; original value) and 113~G (green; 90\% of the original value). \label{fig:sim3comparisonB}}
\end{figure*}

Along with the histograms of the observed signals, Fig.~\ref{fig:sim2comparisonB} also displays the histograms of the polarization signals predicted at $\Theta=10^{\circ}$ by simulation \#2  (red curves), where the average value of the magnetic field is $\langle B(\tau=1) \rangle =149$~G (see Sect.~\ref{sec:simulations}). Likewise, Fig.~\ref{fig:sim3comparisonB} shows (also in red) the predicted histograms for the polarization signals at $\Theta=10^{\circ}$, but for simulation \#3 where the average value of the magnetic field is $\langle B(\tau=1) \rangle =125$~G.

As it can be seen, simulations \# 2 and \#3 with these mean magnetic field strengths both predict excessively strong polarization signals compared to the observations. Therefore, we tried to reduce the magnetic field by a constant factor to see if this would result in histograms closer to the observed ones. The idea is that a reduction of the magnetic field by a constant factor will not have a large effect on how the thermodynamic and kinematic parameters of the simulation behave, and therefore we can keep those to calculate the synthetic profiles without having to run new MHD simulations. This is supported by the fact that the magnetic field is, in all cases, dynamically weak (i.e., below the equipartition value: $B \leq \sqrt{8 \pi P_g} \approx 500 - 1000 \ \SI{}{\gauss,}$ where $P_g$ is the gas pressure). The magnetic field in simulations \#2 and \#3 was reduced from its original values of $\langle B \rangle = \SI{149}{\gauss}$ and $\langle B \rangle = \SI{125}{\gauss}$, respectively (see Table~\ref{tab:B}), to the values shown in Table ~\ref{tab:reducedB}. We note that we reduced the total magnetic field instead of only one of its components (i.e., $B_z$) in order to keep the distribution of the inclination of the magnetic field the same as in the original simulation (see Fig.~\ref{fig:sim_inclination}). Figures~\ref{fig:sim2comparisonB} and ~\ref{fig:sim3comparisonB} include, in blue color, the predicted histograms for one of those reduction factors, namely a factor $0.7$, meaning that $\langle B(\tau=1) \rangle =105$~G in simulation \# 2, and a factor $0.9$ in simulation \# 3, meaning that $\langle B(\tau=1) \rangle =113$~G.

\begin{table}[h]
    \centering
    \caption{Factors used to reduce the average magnetic field in the simulations}
    \begin{tabular}{c c c c }
    \hline
        factor & $\langle B \rangle$ [$\SI{}{\gauss}$] of sim2 & $\chi^2$ of V& $\chi^2$ of L\\
    \hline
        $1.0$ & $149$ & $0.268$ & $0.185$\\
        $0.9$ & $135$ & $0.206$ & $0.183$\\
        $0.8$ & $120$ & $0.165$ & $0.184$\\
        $0.7$ & $105$ & $0.143$ & $0.218$\\
        $0.6$ & $90$ & $0.157$ & $0.249$\\     
    \hline
        factor & $\langle B \rangle$ [$\SI{}{\gauss}$] of sim3 & $\chi^2$ of V& $\chi^2$ of L\\
    \hline
        $1.0$ & $125$ & $0.133$ & $0.085$\\
        $0.9$ & $113$ & $0.071$ & $0.096$\\
        $0.8$ & $100$ & $0.041$ & $0.189$\\   
    \end{tabular}
    \tablefoot{Simulations \# 2 and \# 3 were multiplied with factors in the first column to get lower values for the average magnetic field (second column). The third and fourth columns give the respective $\chi^2$ value for the linear and circular polarization when compared to the observations at $\Theta = 10^\circ$}
    \label{tab:reducedB}
\end{table}


\subsection{Simulation \#1}
\label{sec::discussion sim2}

Figure \ref{fig:discussion1} shows that at $\Theta=0^{\circ}$, simulation \# 1  (orange lines) yields histograms for the polarization signals that are relatively close to the observations. There is still a slight excess in circular polarization at values of $\max\|V(\lambda)\| \ge 10^{-2}$. This suggests that simulation \#1 features too many pixels with vertical magnetic fields. This is also supported by the fact that there seems to be a lack of linear polarization signal in the range $[3 \cdot 10^{-3}$, $8 \cdot 10^{-3}]$. Therefore, despite being already rather horizontal (see green curve in Fig.~\ref{fig:sim_inclination}), the magnetic field should be even more inclined ($\gamma \rightarrow 90^{\circ}$) in order to produce a better fit to the observed histograms. However, overall, it seems as if the magnetic field of this simulation is close to the real magnetic field in the IN regions of the Sun.

At larger heliocentric angles (see Figs.~\ref{fig:discussion1}-\ref{fig:discussion2}), the mismatch between the observed and simulated circular polarization increases, with simulation \# 1 overestimating the amplitude of the observed $V$ signals by an increasing amount as $\Theta$ rises. Meanwhile, the simulated linear polarization gets closer to the observed one up to $\Theta = \SI{40}{\degree}$ but shows an excess in the predicted $[Q, U]$ signals at $\Theta = \SI{50}{\degree}$ and $\Theta = \SI{60}{\degree}$.

This might be explained by the fact that, in the observations, some pixels harboring mostly vertical fields (and thus appearing in the circular polarization when looking from lower $\Theta$) appear in linear polarization when looking at higher $\Theta$, thereby increasing the number of signals in the latter while lowering the signals in the former. While this effect can be seen mostly in the observations, simulation \# 1 also features this decrease in circular polarization for higher viewing angles, albeit in a less pronounced fashion. As the $\tau = 1$ layer is higher in the atmosphere for higher viewing angles (see Fig.~\ref{fig:zoftau}), one can conclude that at this height there is too much magnetic field parallel to the observer's line of sight ($B_\parallel$), while there is seemingly enough magnetic field perpendicular to the observer's line of sight ($B_\perp$). Relating this back to the nonrotated point of reference, this can be interpreted as the magnetic field in simulation \# 1 possessing too many pixels with mostly horizontal fields when sampling higher atmospheric layers.

 This suggests that, in the real IN, the $B_\perp/B_\parallel$ ratio increases more slowly with height, similarly to that predicted by the local dynamo simulations (see Fig.~1 in \citet{steiner_simul} and Fig.~3 in \citet{manfred_simul}). However, this conclusion can only be made for a very limited range of heights of about 40-50~km, which corresponds to the height difference being sampled at $\Theta=0^{\circ}$ and at $\Theta=60^{\circ}$.
 
\subsection{Simulation \#2}
\label{sec::discussion sim1}

In Fig.~\ref{fig:sim2comparisonB}, we see that reducing the magnetic field by a factor of $0.7$ moves the histogram (right of the noise-peak) down without changing the slope, meaning that the power law in the predicted polarization signals remains the same regardless of the mean field strength. It is therefore difficult to assess at which $\langle B \rangle$ the simulation best
fits the observations, as it depends on the polarization level at which one wants the histograms to agree. Overall, out of all values in Table~\ref{tab:reducedB}, we choose $\langle B \rangle = \SI{105}{\gauss}$ (blue curves), as it seems to produce the{ lowest $\chi^2$ value in the circular polarization}. However, even after the reduction of the magnetic field, simulation \# 2 does not produce similar histograms to the observations at $\Theta = \SI{0}{\degree}$.
Also, when looking at $\Theta>\SI{0}{\degree}$ (see Figs.~\ref{fig:discussion1}-\ref{fig:discussion2}), the slope of the circular polarization still does not come close to the observations, meaning that at those heights the power law of the distributions is still not the same. The linear polarization also seems to follow a different power law than the observations at $\Theta= \SI{0}{\degree}$ to $\Theta=\SI{20}{\degree}$ (blue curve in Fig.~\ref{fig:discussion1}), while the overall linear polarization beyond the noise peak is underestimated in the simulation (with reduced magnetic field) until $\Theta=\SI{30}{\degree}$ and is overestimated afterward.

\subsection{Simulation \#3}
\label{sec::discussion sim3}

The simulation with horizontal fields, for a reduction factor of $0.9$ and thus $\langle B(\tau=1) \rangle = 113$~G, fits the observed histograms at $\Theta=0^{\circ}$ reasonably well: compare black and green curves in Fig.~\ref{fig:sim3comparisonB} { (even though the $\chi^2$ for the circular polarization would still decrease for a lower factor, this factor was chosen to mitigate the increase in $\chi^2$ in the linear polarization)}. In Fig.~\ref{fig:discussion1}, we see that, at $\Theta = \SI{0}{\degree}$ (top right panel), the predicted linear polarization (green curve) has a similar distribution to the observed signals (black curve) at this heliocentric angle, while the circular { polarization} (top left panel) has a surplus between $4 \times 10^{-3}$ and $0.01$. At the same time, the observations show a greater number of pixels with strong polarizations, that is of $\ge 0.03$. The trend of the surplus of circular polarization keeps increasing with higher viewing angles (see also Fig.~\ref{fig:discussion2}). At the same time, the strong circular polarization signals sometimes seem to match those of the observations ($\Theta = \SI{50}{\degree}$; the green curve in the middle row in Fig.~\ref{fig:discussion2}), whereas at other times, there seems to be a shortage (i.e. $\Theta = \SI{30}{\degree}$; the green curve in bottom row in Fig.~\ref{fig:discussion1}). The linear polarization shows a shortage after $\SI{30}{\degree}$ compared to the observations.

It is important to note that part of the disagreement between observations and simulated histograms for large circular polarization signals above 0.03 and also for the linear polarization signals after the noise-peak for viewing angles of larger than $\Theta = \SI{30}{\degree}$ can be caused by the low statistics in the simulations, which becomes particularly pressing after degradation and resampling (see Sect.~\ref{sec:degradation}). This is particularly the case in CO5BOLD simulations (simulation \# 3), because it features a significantly smaller domain than MURaM, and where only$\sim 3400$ pixels {remain after the resampling (Sect.~\ref{sec:degradation}) for the creation of the histograms} that are compared to the observations. However, in the case of the weaker circular polarization signal, it is clear that simulation \# 3 predicts too much of it. In conclusion, although we cannot really distinguish exactly what simulation \# 3 is missing that would enable it to better fit the observations, we can state that at least two factors play an important role: not only is a lower field strength needed, but it probably also requires a different distribution of inclinations of the magnetic field ($\gamma$; see Fig.~\ref{fig:sim_inclination}).

\subsection{Effect of adding $30$~G vertical field to SSD}

\begin{figure*}
    \centering
    \includegraphics[height=0.3\textwidth]{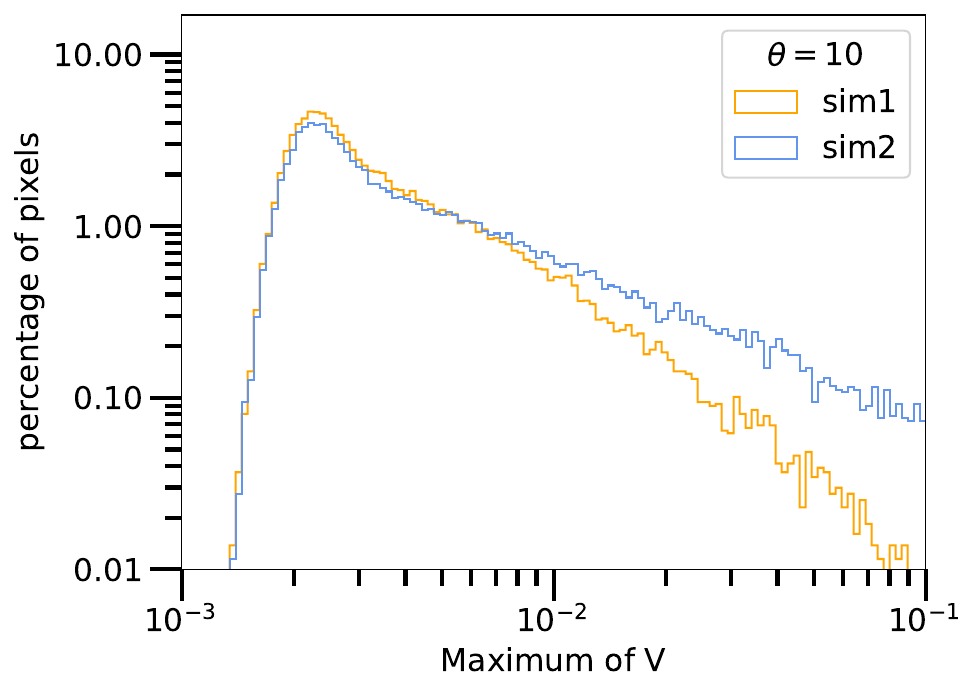}
    \includegraphics[height=0.3\textwidth]{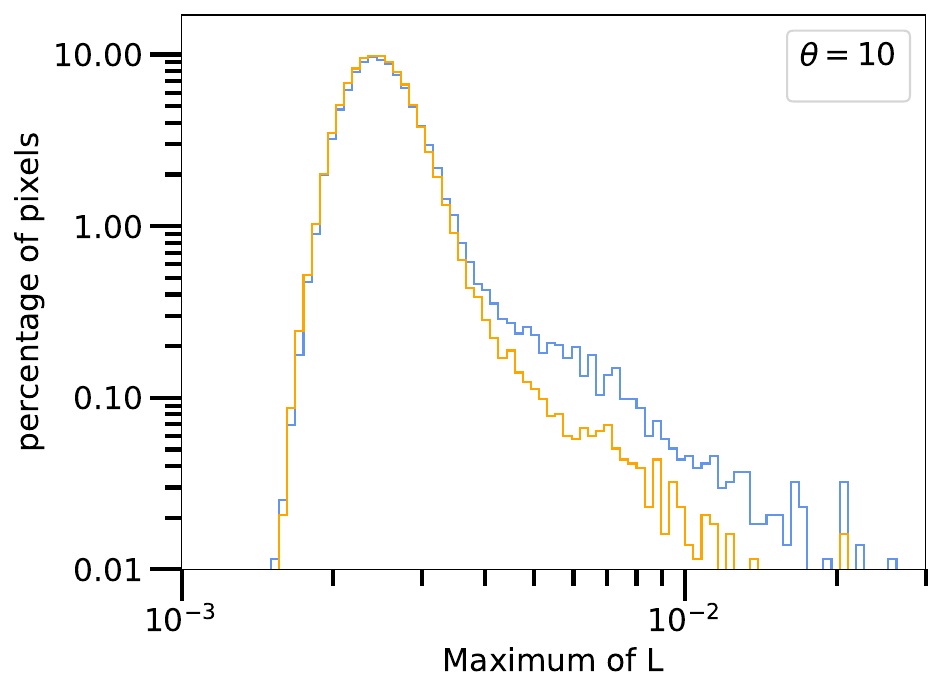}
    \caption{Comparison of the circular and linear polarization between the small-scale dynamo simulation (simulation \#1) and a small-scale dynamo simulation (simulation \#2) where a $30$~G vertical field is added.\label{fig:sim1and2}}
\end{figure*}

As described in Sect.~\ref{sec:simulations}, simulation \# 1 is an SSD simulation, whereas simulation \# 2 was evolved for 6 hours starting from a similar SSD simulation where a vertical field of $B_z = 30$~G 
was added over the entire domain. It is therefore interesting to study the effect of adding this vertical field  on the predicted histograms for the polarization signals. This effect is visualized in{ Fig.~\ref{fig:sim1and2}}, where we present the predicted circular (left panel) and linear (right panel) polarization signals for: simulation \# 1 (orange) and \# 2 (blue). We note that, in these histograms, the original values of the magnetic field have been used (see Table~\ref{tab:B}) and the network has not been removed, allowing us to see both low and high polarization signals.

As demonstrated by Fig.~\ref{fig:sim1and2} adding 30~Gauss of vertical field modifies the histograms both in the region of high ($\ge 5\times10^{-2}$) and low ($\in [2\times 10^{-3},5\times 10^{-2}]$) polarization signals. This suggests that the vertical field evolves in such a way that its affects both the network and the IN. The latter can indeed be affected by the inclusion of a vertical field, as this suppresses convection and thus also the feedback mechanism from the velocity field into the magnetic field via induction (i.e., the so-called dynamo).

\section{Conclusions}

We confronted synthesized polarimetric spectra of three simulations, each representing a different scenario for the origin of the magnetic field in the solar internetwork (IN), to Hinode observations of the quiet sun by statistically comparing the circular and linear polarization signals. Simulation \#1 has a small seed magnetic field that the local dynamo amplifies until saturation, while simulation \#2 has an initial vertical magnetic field imposed on top of another local dynamo simulation. Finally, simulation \#3 features horizontal fields advected from the bottom boundary. From these simulation snapshots, we synthesized the Stokes vector for different viewing angles $\Theta$ (Sect.~\ref{sec:synthesis}). We then degraded the synthesized spectra with the spectral and spatial PSF of Hinode and resampled the grid size to match the Hinode data (Sect.~\ref{sec:degradation}). Finally, we added photon noise to the simulated data to make the synthesized spectra comparable to the observations. The last step consisted of removing the pixels that belong to the network (Sect.~\ref{sec:: removal of network})
from both Hinode observations and synthesized spectra.

For simulations \#2 and \#3, we found that the polarization signals were too strong compared to the observations (see Fig.~\ref{fig:sim2comparisonB} and Fig.~\ref{fig:sim3comparisonB}), and so we lowered the magnetic field in the simulation snapshots by multiplying it by a constant factor and synthesized the spectra again. Once these corrections are performed, we find that simulation \# 2, which features an initial vertical field, { yields $\chi^2$ values with the observations that are a factor of 2-10 larger than in the other simulations. We conclude therefore that simulation \# 2} does not correctly represent the physical conditions in the solar IN, either in terms of the total field strength or in the inclination of the magnetic field. We can also conclude that our results
do not support the scenario where these vertical fields originate from the decay or recycling of active regions \citep{originDecay}.
{ Simulation \# 3, with initial horizontal fields at the bottom boundary, produces histograms of linear polarization (${\rm max} [\|Q(\lambda)\|,\|U(\lambda)\|]$) that most closely resemble those derived from the observations, with $\chi ^2$ values of 20 to 50 \% lower than the other simulations. However, in the circular polarization, the $\chi^2$ values are higher by a factor of 2 to 3 than those found for the first simulation. This suggests that the emergence of horizontal fields from the convection zone, as suggested by the global dynamo \citep{originDynamolarge}, is unlikely to be the origin of the magnetic fields in the solar IN.}

Simulation \# 1, representing a small-scale dynamo occurring locally in the photosphere, produces the best fits{ overall} to the observed histograms.{ Although $\chi^2$ values in the linear polarization are slightly higher than in simulation \# 3, $\chi^2$ values in the circular polarization are significantly smaller} at all studied heliocentric angles. This leads us to conclude that, of all three possible scenarios studied in this work, this 
one is{ the most likely to explain} the origin of the IN. Similar conclusions were reached by \citet{Danilovic2010} and \citet{Lagg2016IR}, who performed comparisons between small-scale dynamo simulations and quiet Sun observations using Hinode/SOT/SP or the GRIS instrument on the GREGOR solar telescope. However, the agreement with the observations is still not excellent for the circular polarization at large heliocentric angles. We surmise that this simulation would produce an even better agreement with the observations if the ratio between the horizontal and vertical components of the magnetic field in simulation \# 1 were not to grow so quickly with height.

As our work is mostly limited by the fact that only three different simulations were used, it would be desirable to improve our analysis by{ looking at more} simulations with different initial conditions. Another possibility would be to employ other spectral lines to see deeper or higher in the solar atmosphere so as to better constrain how the magnetic field changes with height.

\begin{acknowledgement}
We gratefully acknowledge Ostar Steiner, Flavio Calvo, Sanja Danilovic, and Matthias Rempel for their help with CO5BOLD and MURaM MHD simulations and useful comments on the paper. Many thanks also to Bruce W. Lites for many useful discussions throughout the years about the internetwork magnetic fields.
This research has made use of NASA's Astrophysics Data System. Hinode is a Japanese mission developed and launched by ISAS/JAXA, collaborating with NAOJ as a domestic partner, and NASA and STFC (UK) as international partners. Scientific operation of the Hinode mission is conducted by the Hinode science team organized at ISAS/JAXA. This team mainly consists of scientists from institutes in 
the partner countries. Support for the post-launch operation is provided by JAXA and NAOJ (Japan), STFC (U.K.), NASA, ESA, and NSC (Norway). 
\end{acknowledgement}
\bibliography{qs_paper_corr}

\begin{appendix}
\section{Additional figures}
\begin{figure*}
    \centering
    \includegraphics[height=0.3\textwidth,width=0.449\textwidth]{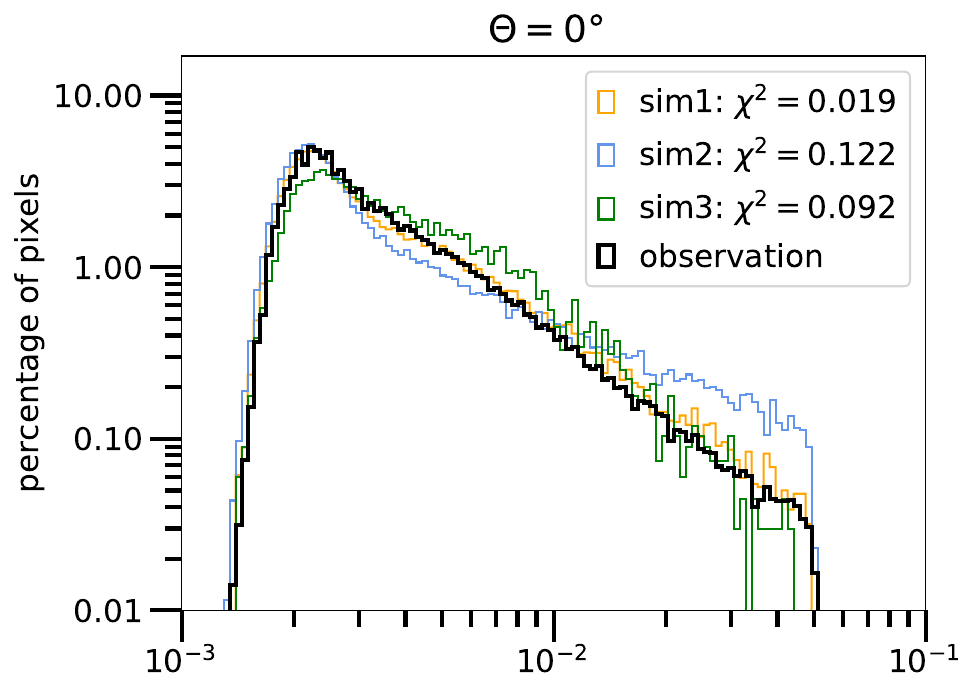}
    \includegraphics[height=0.3\textwidth,width=0.414\textwidth]{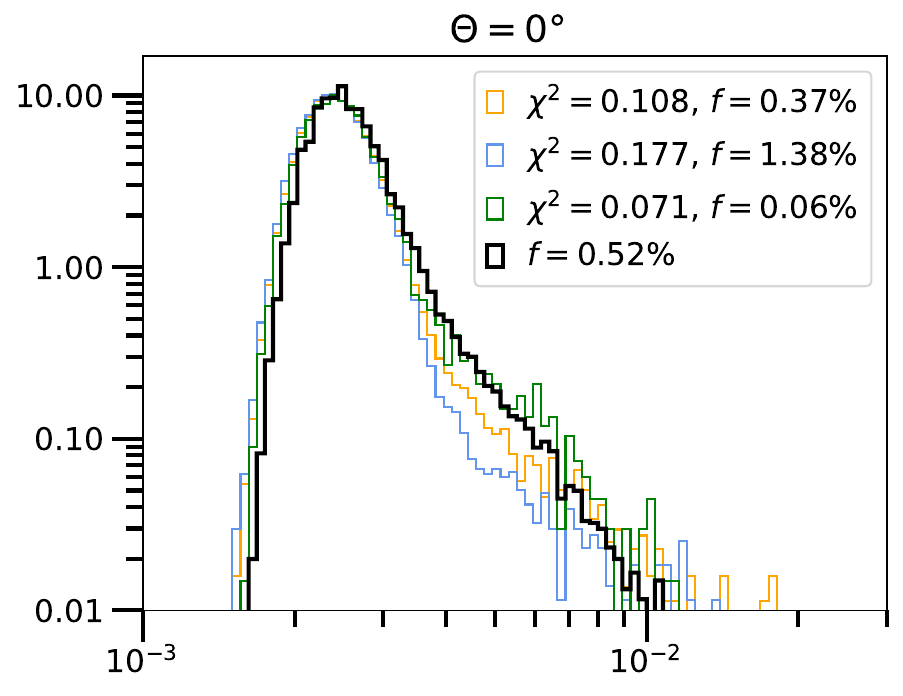}
    \includegraphics[height=0.3\textwidth,width=0.449\textwidth]{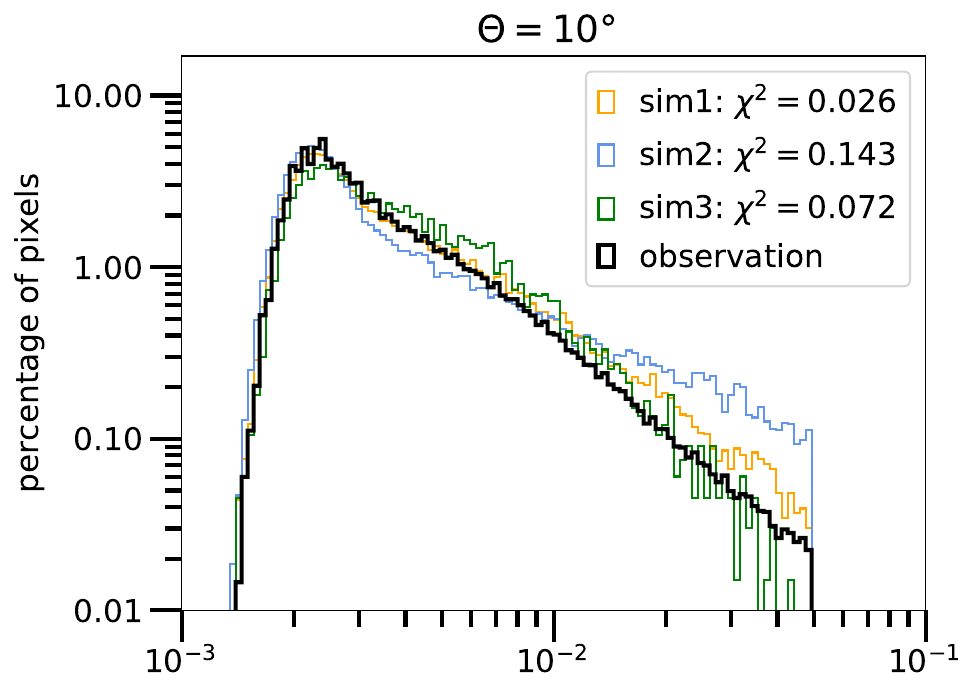}
    \includegraphics[height=0.3\textwidth,width=0.414\textwidth]{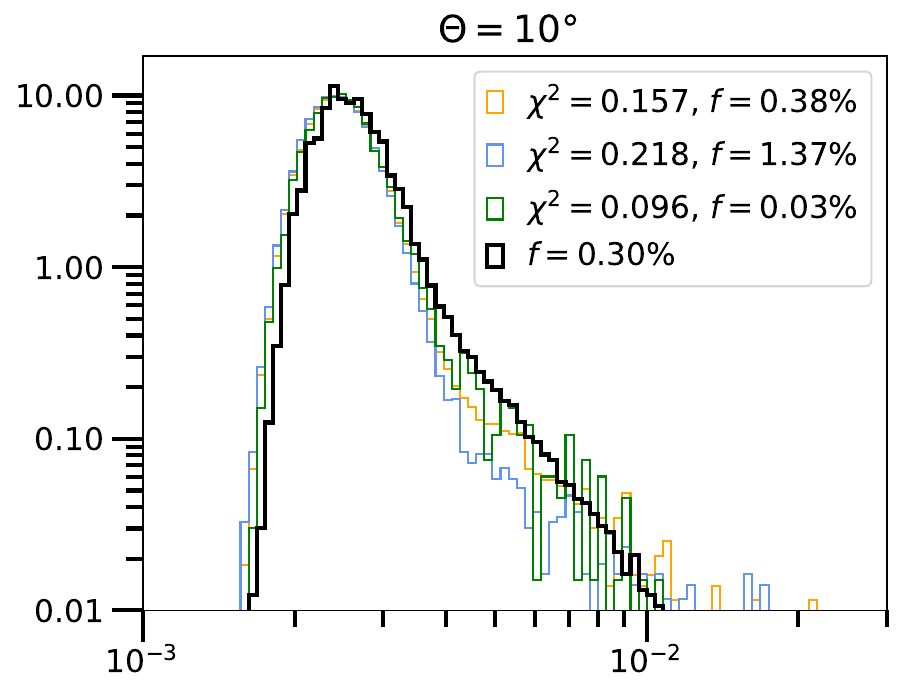}
    \includegraphics[height=0.319\textwidth,width=0.449\textwidth]{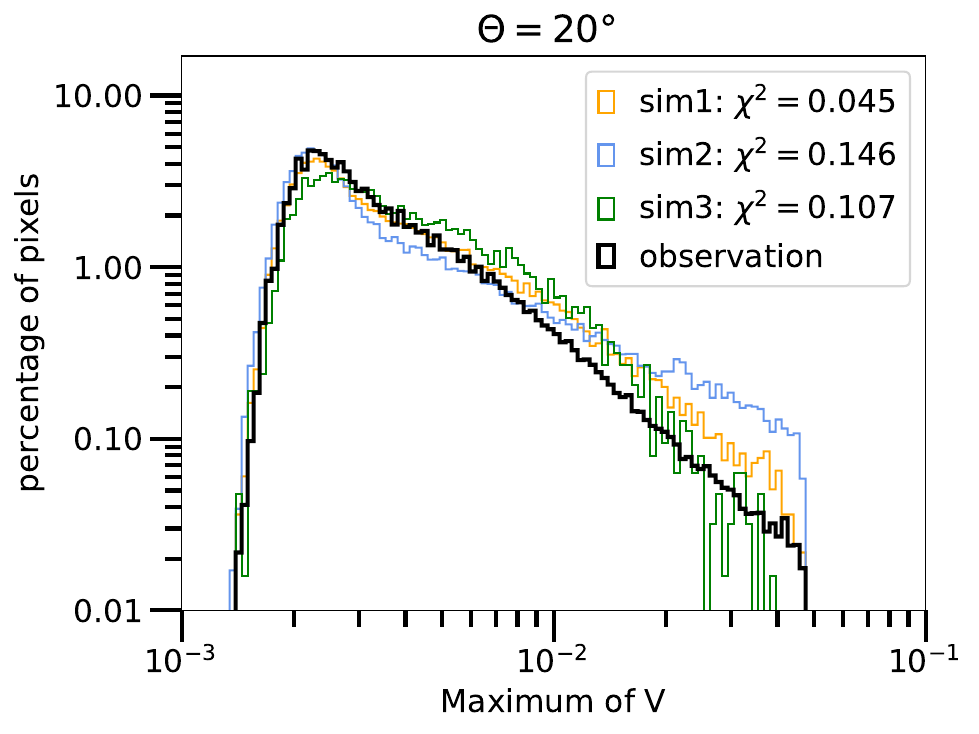}
    \includegraphics[height=0.319\textwidth,width=0.414\textwidth]{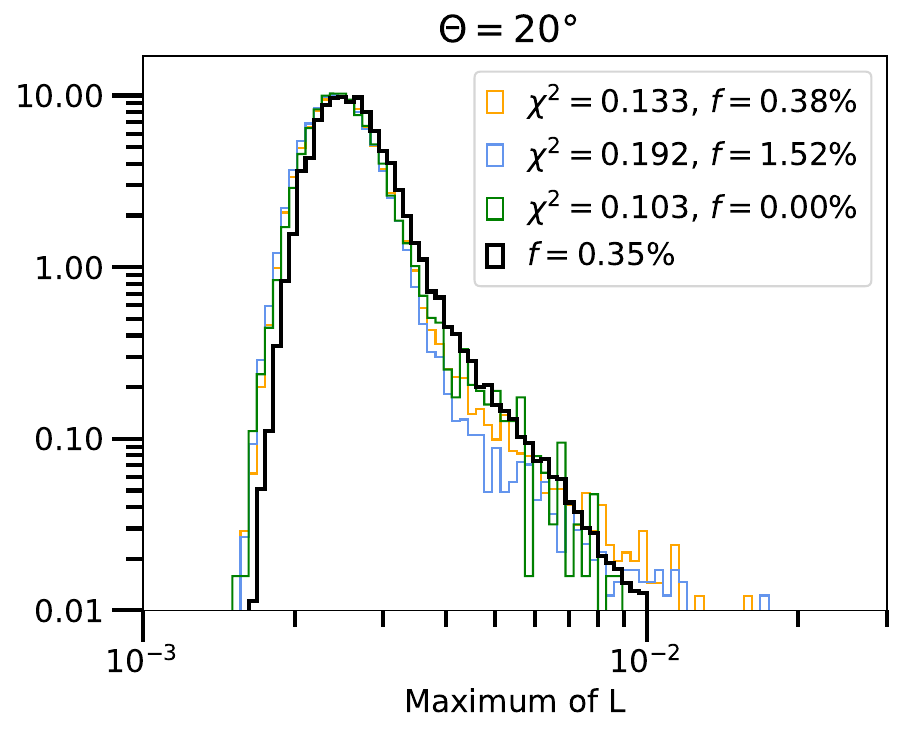}
    \caption{Histogram of the number of pixels as a function of the maximum value in their polarization signals -- circular polarization (left) and linear polarization (right) -- normalized to the average continuum intensity for viewing angles of $\Theta = \SI{0}{\degree}$ (top), $\Theta = \SI{10}{\degree}$ (middle), and $\Theta = \SI{20}{\degree}$ (bottom). { The legend shows the $\chi^2$ between the simulation and observations for each polarization. It also shows the fraction, $f$, of pixels that are removed from the analysis for being considered part of the network.}\label{fig:discussion1}}
\end{figure*}
\begin{figure*}
    \centering
    \includegraphics[height=0.3\textwidth,width=0.449\textwidth]{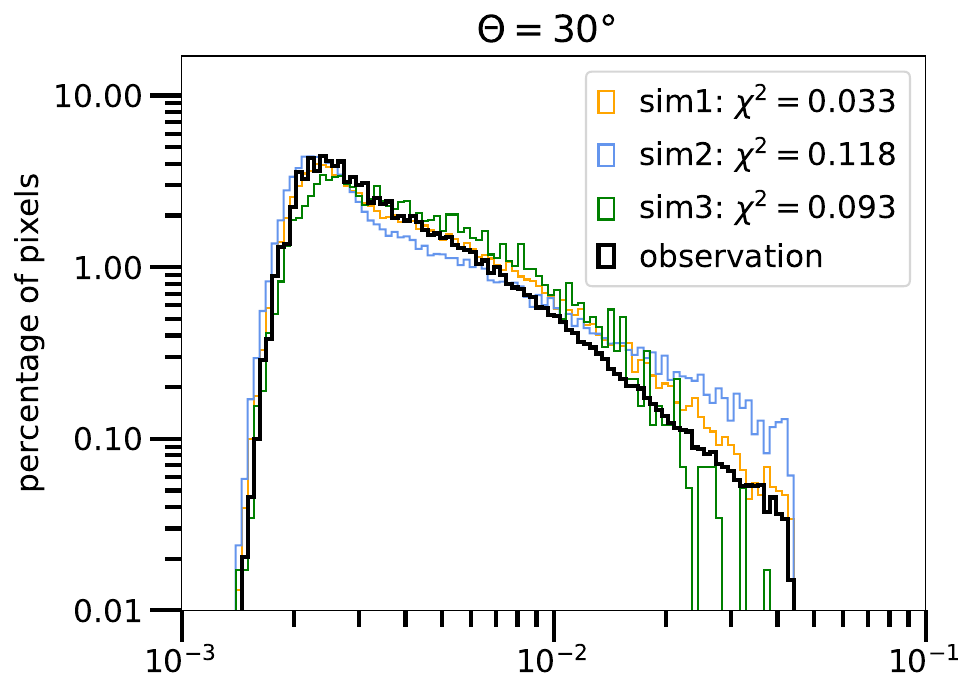}
    \includegraphics[height=0.3\textwidth,width=0.414\textwidth]{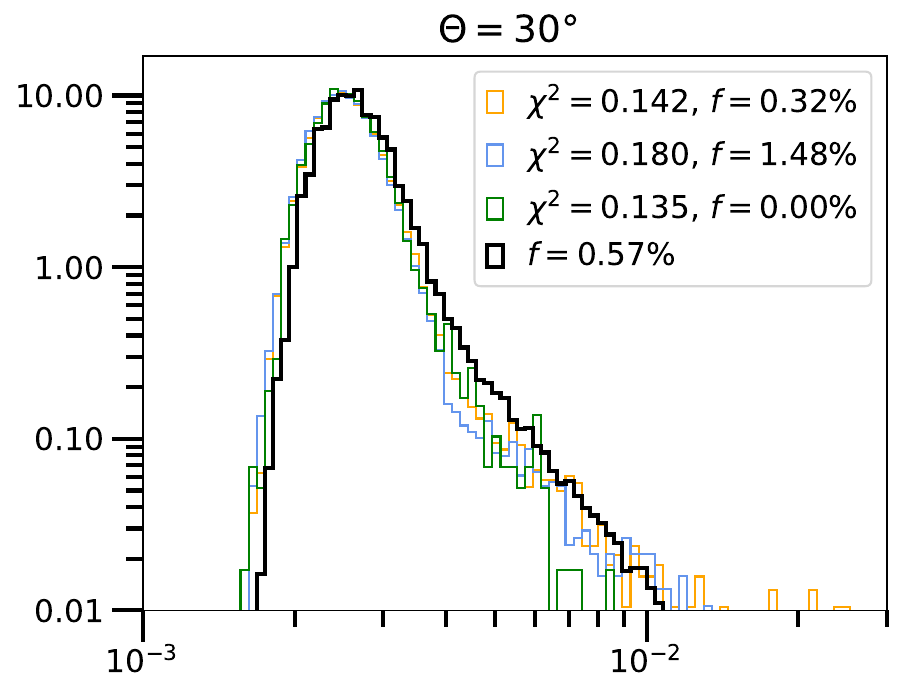}
    \includegraphics[height=0.3\textwidth,width=0.449\textwidth]{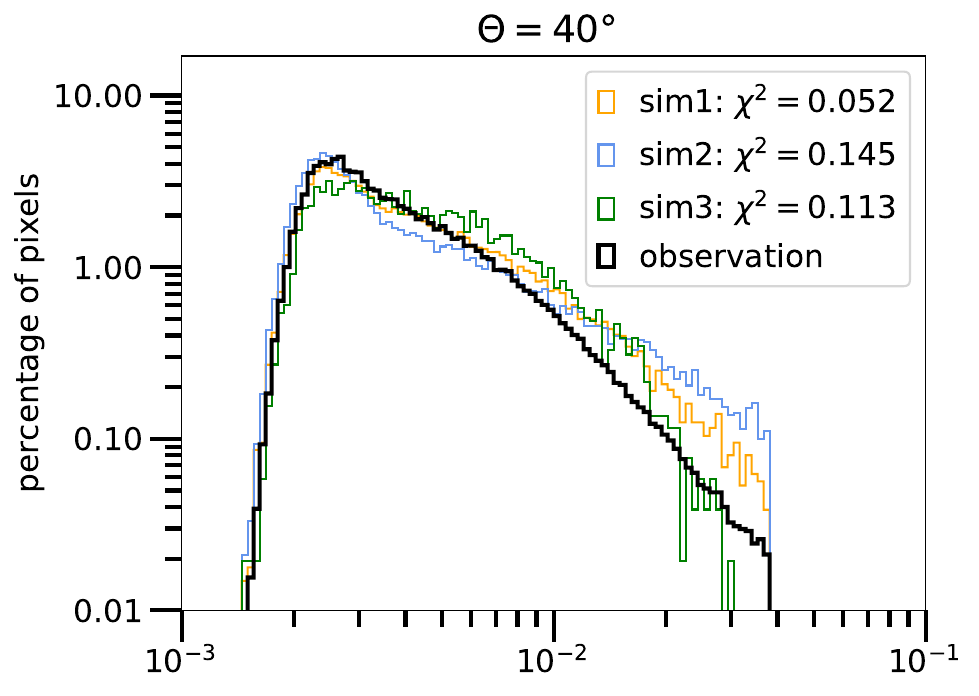}
    \includegraphics[height=0.3\textwidth,width=0.414\textwidth]{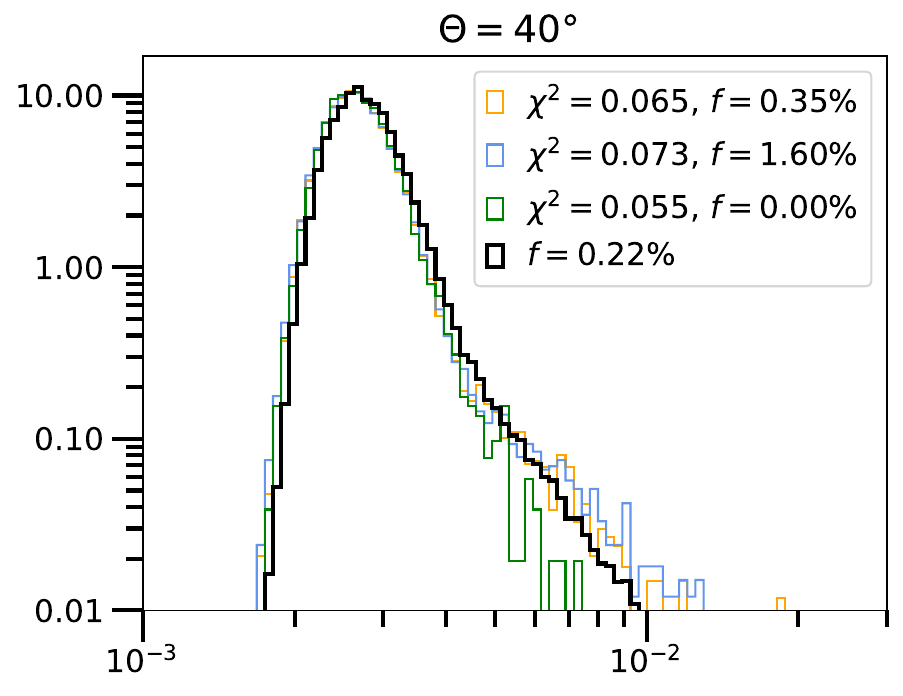}
    \includegraphics[height=0.3\textwidth,width=0.449\textwidth]{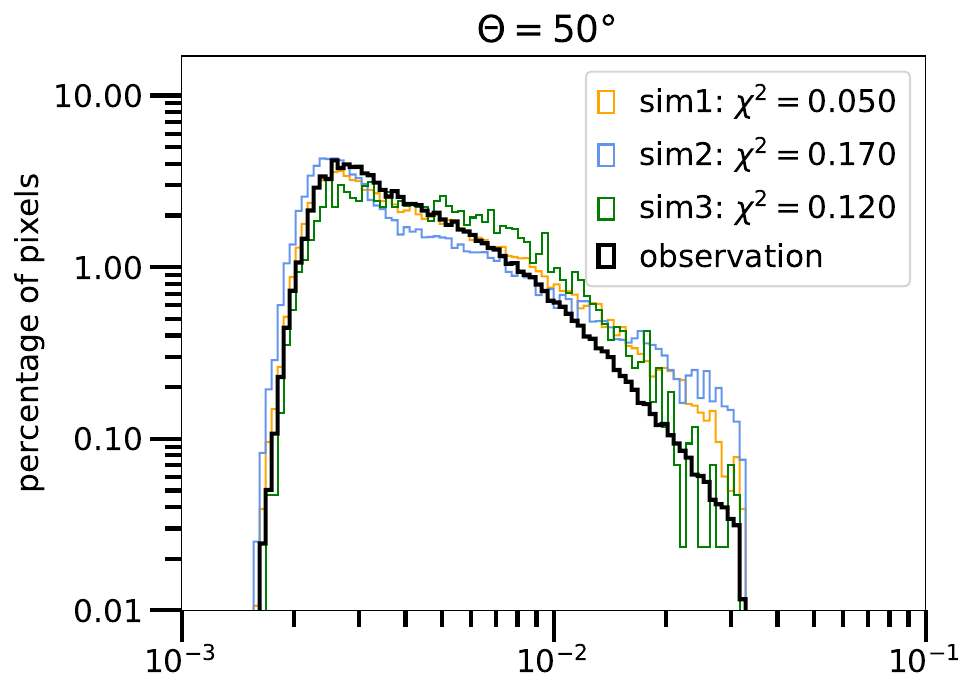}
    \includegraphics[height=0.3\textwidth,width=0.414\textwidth]{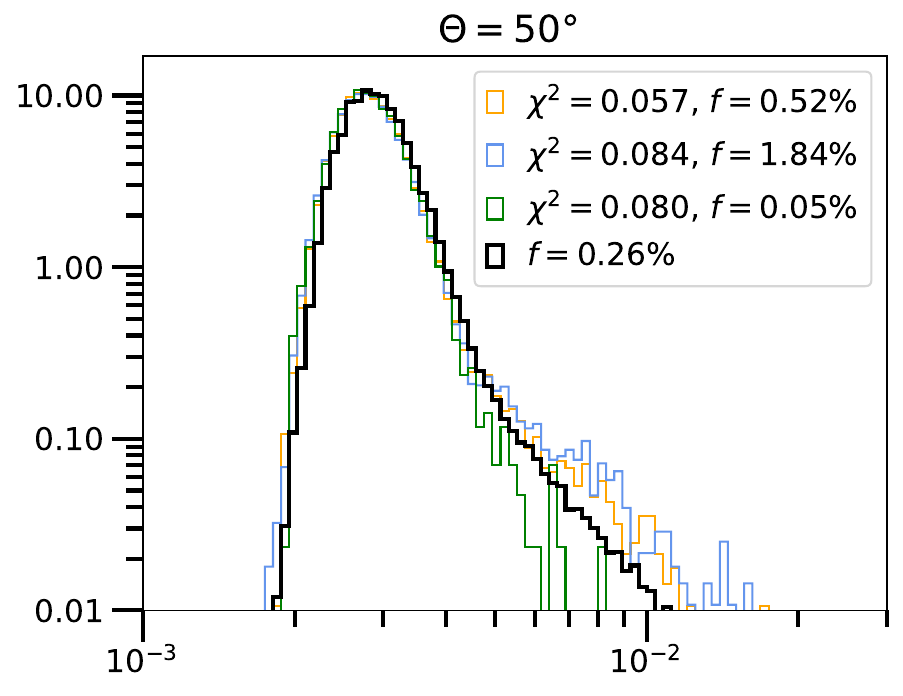}
    \includegraphics[height=0.319\textwidth,width=0.449\textwidth]{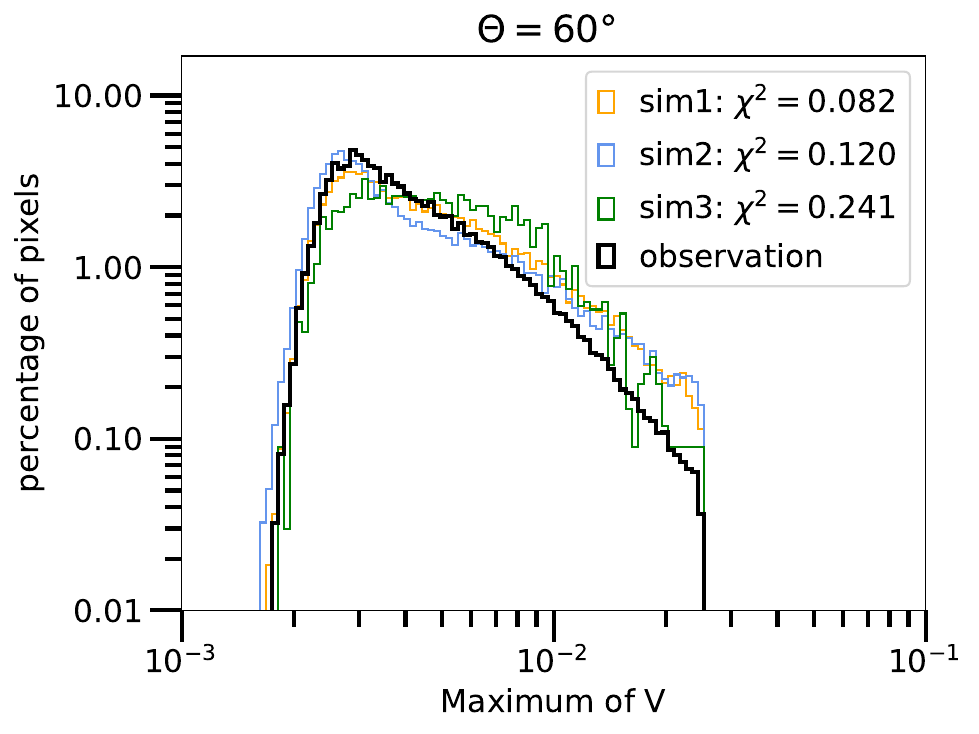}
    \includegraphics[height=0.319\textwidth,width=0.414\textwidth]{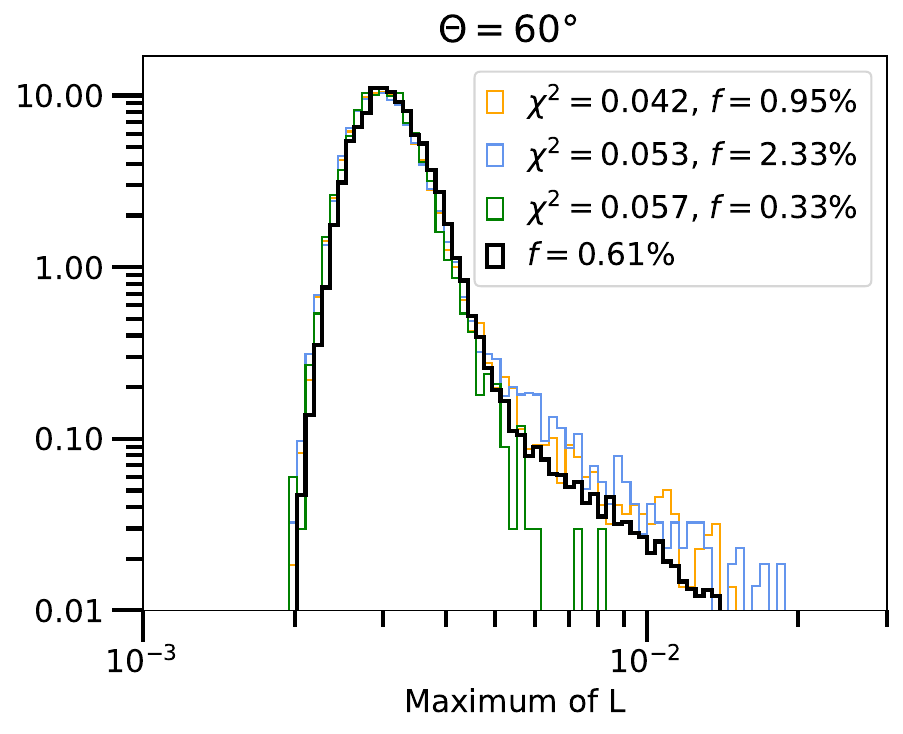}
    \caption{Same as Fig.~\ref{fig:discussion1} but for viewing angles of $\Theta = \SI{30}{\degree}$ (top), $\Theta = \SI{40}{\degree}$ (middle top), $\Theta = \SI{50}{\degree}$ (middle bottom), and $\Theta = \SI{60}{\degree}$ (bottom).\label{fig:discussion2}}
\end{figure*}

\end{appendix}


\end{document}